\RequirePackage{fix-cm}

\documentclass[smallextended]{svjour3}       
\smartqed 
\usepackage{graphicx}
\usepackage{amsmath}
\usepackage{amssymb}
\usepackage{bbm}
\usepackage{cite}
\usepackage[colorlinks=true,citecolor=blue,urlcolor=blue,breaklinks]{hyperref}
\usepackage{latexsym}
\usepackage[normalem]{ulem}
\usepackage{pdfsync}
\newcommand{\MYhref}[3][blue]{\href{#2}{\color{#1}{#3}}}%

\usepackage{xcolor}

\begin{document}

\title{Line-Intensity Mapping: Theory Review}
\subtitle{with a focus on star-formation lines} 

\author{J.~L.~Bernal \and
         E.~D.~Kovetz
}

\institute{J.~L.~Bernal \at
            William H. Miller III Department of Physics and Astronomy, Johns Hopkins University\\
	Baltimore, MD 21218, USA\\
	\email{jbernal2@jhu.edu}  \and
	E.~D.~Kovetz \at
              Department of Physics, Ben-Gurion University of the Negev\\
              Be'er Sheva 84105, Israel\\
              \email{kovetz@bgu.ac.il}}

\date{Received: date / Accepted: date}

\maketitle

\begin{abstract}

Line-intensity mapping (LIM) is an emerging approach to 
survey the Universe, using relatively low-aperture instruments to scan large portions of the sky and collect the total spectral-line emission from galaxies and the intergalactic medium.
Mapping the intensity fluctuations  
of an array of lines 
offers a unique opportunity to probe redshifts well beyond the reach of other cosmological observations, access regimes that cannot be explored otherwise, and exploit the enormous potential of cross-correlations with other measurements. 
This promises to deepen our understanding of  
various 
questions related to galaxy formation and evolution, cosmology, and fundamental physics. 

Here we focus on lines ranging from microwave to  optical frequencies, the emission of which is related to star formation in galaxies across cosmic history.
Over the next decade, LIM will transition from a pathfinder era of first detections to an early-science era where data from more than a dozen missions will be harvested to yield new insights and discoveries. This review 
discusses the primary target lines for these missions, describes the different approaches to modeling  
their intensities and fluctuations, surveys the scientific prospects of their measurement, presents the formalism behind the statistical methods to analyze the data, and motivates the opportunities for synergy with other observables. Our goal is to provide a pedagogical introduction to the field for non-experts, as well as to serve as a comprehensive reference for specialists.

\end{abstract}

\newpage

\setcounter{tocdepth}{2}

{
  \hypersetup{linkcolor=black}
  \tableofcontents
}

\section{Introduction}
\label{sec:intro}

In the past couple of decades, cosmology has become a precision science, with increasingly detailed measurements of the cosmic microwave background (CMB) and ever-larger surveys of galaxies yielding percent-level constraints on the parameters of the standard model of cosmology, $\Lambda$CDM, and informing sophisticated models of galaxy formation and evolution.  In spite of this success, the nature of key theoretical pillars of $\Lambda$CDM remains elusive, the evolution of galaxy properties is largely unconstrained and several experimental tensions that have arisen warrant explanation.   Looking forward, novel cosmological observables will be crucial to making progress in this quest to deepen our understanding of the cosmos.

Line-intensity mapping (LIM)~\cite{1979MNRAS.188..791H,Suginohara:1998ti,Chang:2007xk, Visbal:2010rz,Visbal:2011ee, Gong:2011ts, Carilli:2011vw, Fonseca:2016qqw, Kovetz:2017agg} measures the integrated emission of spectral lines originating from many individually unresolved galaxies and the diffuse intergalactic medium (IGM). It tracks the makeup and growth of cosmic structure as well as the history of the astrophysical processes controlling galaxy formation and evolution. Unlike galaxy surveys, LIM does not require resolved high-significance detections but uses  all incoming photons from any source within the field of view, obtaining tomographic line-of-sight information from targeting a known spectral line at different frequencies. This enables the use of smaller low-aperture instruments with modest experimental budgets to rapidly survey large areas of  sky for studying cosmology and large-scale astrophysics. Through its intrinsic dependence on cosmology and astrophysics, LIM connects the smallest and largest scales in the Universe as no other observable can do. 

There has been a flurry of interest in LIM in recent years with an impressive lineup of experimental projects pursuing an array of atomic and molecular spectral lines across the electromagnetic spectrum, from the radio to the ultraviolet. Simultaneously, the theoretical framework of the modeling of line intensities and 
the study of the line-intensity maps have vastly developed.

The focus of this review will be on spectral lines associated with gas cooling and star formation in galaxies, ranging from the rotational carbon-monoxide (CO) transitions observed in the sub-mm, through  bright fine-structure lines such as [CII] in the far-infrared, to the hydrogen H$\alpha$ and Ly$\alpha$ lines in the optical and ultraviolet. The HI 21-cm line originating from the neutral hydrogen permeating the intergalactic medium in the early-Universe  and from pockets within galaxies post-reionization has been reviewed
in Refs.~\cite{Furlanetto:2006jb,Morales:2009gs,Pritchard:2011xb,Liu:2019awk}.

As this review will advocate, LIM holds promise to become a cornerstone for our future understanding of cosmology and astrophysics, as the CMB and galaxy surveys have been so far. With just a handful of preliminary detections to date, this declaration may seem premature, but several properties such as its unprecedented reach, unique access, and extended overlap, suggest that LIM offers unique advantages in mapping the large-scale structure in the Universe. 

Circumventing the requirement of individual source detection, LIM avoids the intrinsic depth limitation of surveys of galaxies, which are increasingly faint and sparse at high redshift. There are proposals to extend galaxy surveys to  high redshifts (e.g.\ Ref.~\cite{Schlegel:2019eqc}), but in such regimes LIM may yield  more efficient mapping of cosmological perturbations~\cite{Uzgil:2014pga,Cheng:2018hox, Schaan:2021hhy}, with  markedly lower budgets. 

\begin{figure}[h!]
 \begin{centering}
\hspace{0.9cm} 
\includegraphics[width=\textwidth]{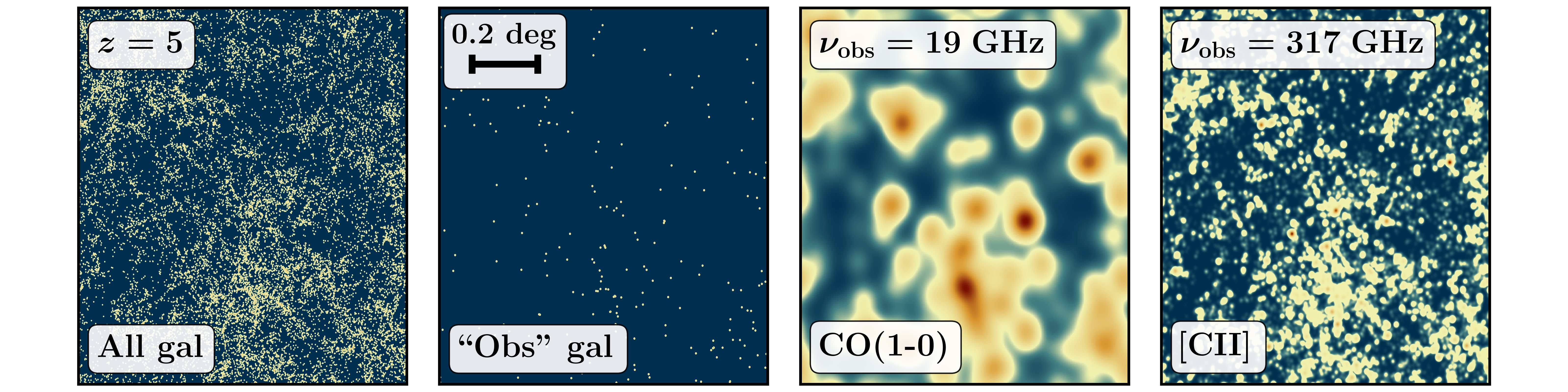}
\caption{A 1 deg$^2$ field at $z\in\left[4.9,5.1\right]$ showing (from left to right) all the galaxies, ``observable'' galaxies (assuming an arbitrary cut off in the stellar masses $\geq 10^{9.5}\,M_\odot$ as a theoretical proxy for detection threshold), and maps of the CO and [CII] intensity fluctuations from this field. Characteristic angular resolutions of instruments targeting each line (4' and 0.5' for CO and CII, respectively) are applied. Results obtained using the approach of Ref.~\cite{LC_paper}, assuming parameterizations from Ref.~\cite{2016ApJ...829...93K} and Ref.~\cite{DeLooze:2014dta}.
}
\label{fig:fields}
\end{centering}
\end{figure}

Figure~\ref{fig:fields} demonstrates the power of LIM in terms of reach. In a simulated volume spanning 1 deg$^2$ on the sky and a redshift slice $z=[4.9,5.1]$, we compare the 
total distribution of galaxies with those that could be realistically observed by a galaxy survey, and show the corresponding emission that would be detected by LIM experiments targeting CO(1-0) and [CII], respectively (not including instrumental noise nor contaminants, to ease the comparison). 
We see that each LIM experiment, with its own resolution limits, has the potential to detect fluctuations even where the galaxy survey would observe a void, with independent line emissions exhibiting different fluctuation amplitudes. While this example helps uncover the potential of LIM to augment and reach beyond existing techniques,  it merely brushes the tip of the iceberg.   

\begin{figure}[h!]
 \begin{centering}
\includegraphics[width=0.9\textwidth]{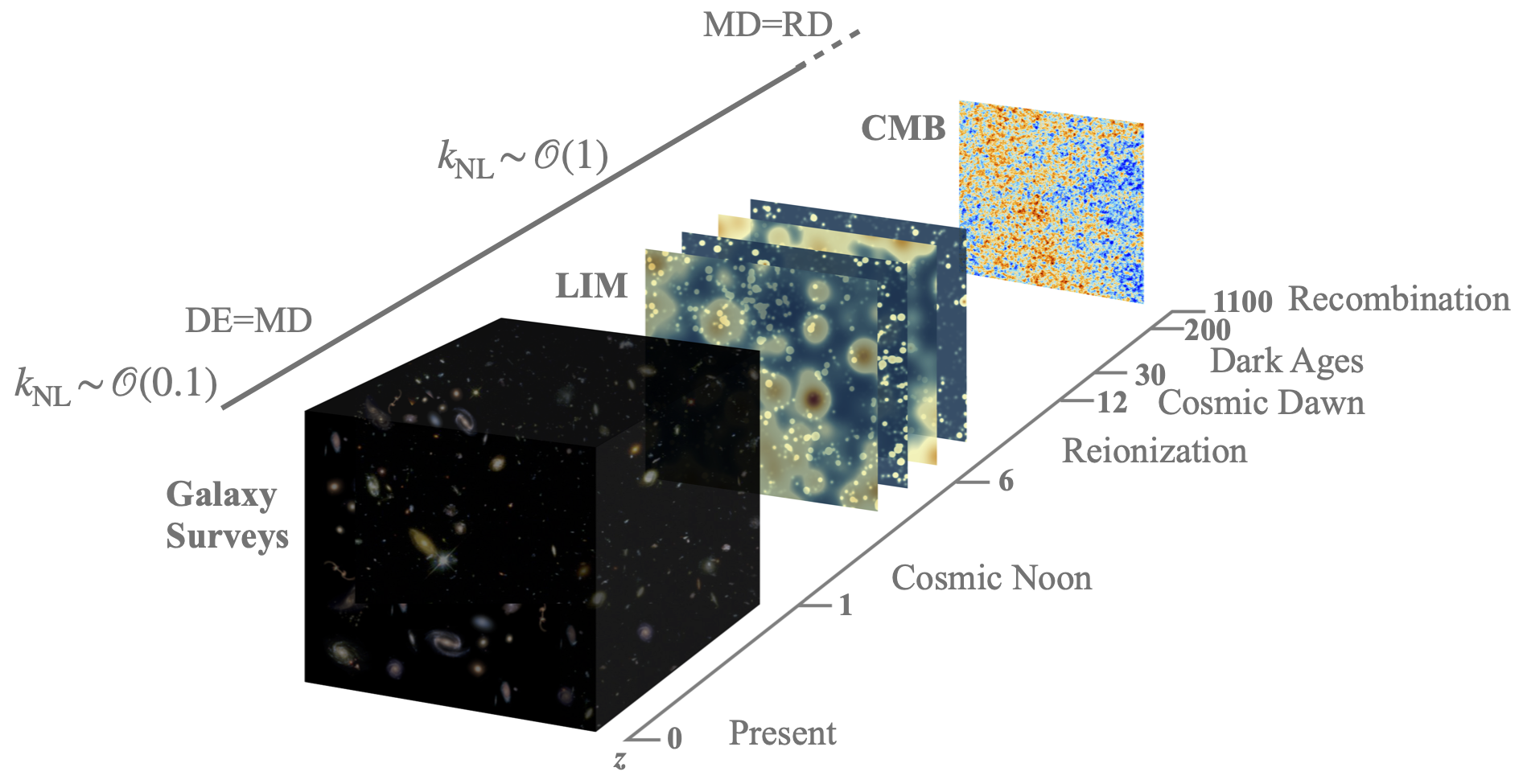}
\caption{Intensity mapping of multiple line emissions  provides rich access to redshift volumes otherwise inaccessible, enabling detailed study of various important epochs in cosmic history. Above the arrow of time on the left, MD=RD and DE=MD denote the redshifts of matter-radiation, and dark energy-matter equality, respectively, and $k_{\rm NL}$ is a rough estimation of the scale at which matter clustering becomes non linear, specified in units of ${\rm Mpc}^{-1}$.} 
\label{fig:epochs}
\end{centering}
\end{figure}

In terms of access, LIM is uniquely poised to probe crucial epochs in the history of our Universe, as illustrated in Fig.~\ref{fig:epochs}. Collecting all incoming photons, LIM directly probes the epoch of reionization (EoR), the IGM, the interstellar medium (ISM) and the formation and makeup of stars, granting access to astrophysical and cosmological information inaccessible otherwise since it is sensitive to the whole population of emitters instead of only the brightest ones. This also makes LIM more robust against selection effects and misestimation of line-emission and luminosity ratios whenever one line has extended emission.

Furthermore, extending the reach of a survey to higher redshifts increases the volume observed dramatically. This allows us to probe larger scales, potentially reaching scales of the order of the horizon, where signatures of inflation 
may be present. Moreover, as redshift grows, non-linear matter clustering is confined to ever-smaller scales, facilitating the theoretical interpretation. Although the access to small scales depends on the experimental resolution and the growing contribution of non-linear bias terms may hinder the ability to extract robust information from 
them~\cite{Desjacques:2016bnm, MoradinezhadDizgah:2021dei}, exploring small scales will significantly increase the constraining power of LIM surveys. Considering different science targets introduces a tradeoff between deeper and wider surveys. 

In terms of overlap, there is tremendous potential for LIM cross correlations. First, multi-line analyses allow for a multi-phase and multi-scale study of galaxy properties~\cite{2019ARA&A..57..511K},
but also for a statistically more powerful study of large-scale structure to probe deviations from $\Lambda$CDM. 
Line-intensity maps can also be  cross-correlated  with galaxy surveys, with CMB observations---to study CMB secondary anisotropies such as weak gravitational lensing and the Sunyaev-Zel'dovich effect---and with catalogs of astrophysical transients such as gravitational waves from merging black holes and fast radio bursts. 

\begin{figure}[h!]
 \begin{centering}
\hspace{0.9cm} 
\includegraphics[width=0.82\textwidth]{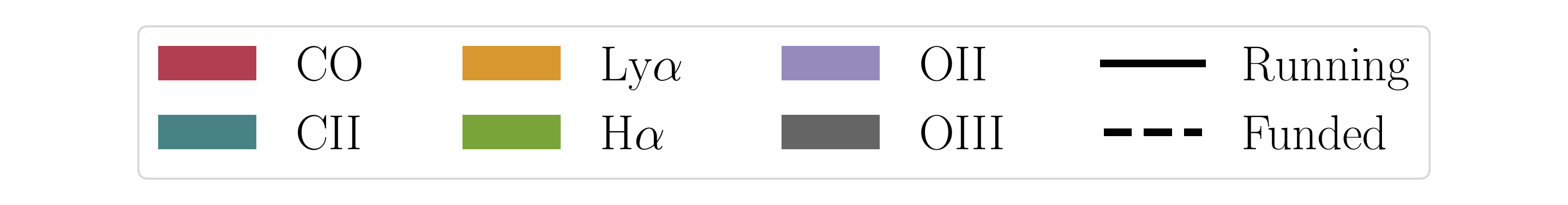}
\includegraphics[width=0.96\textwidth]{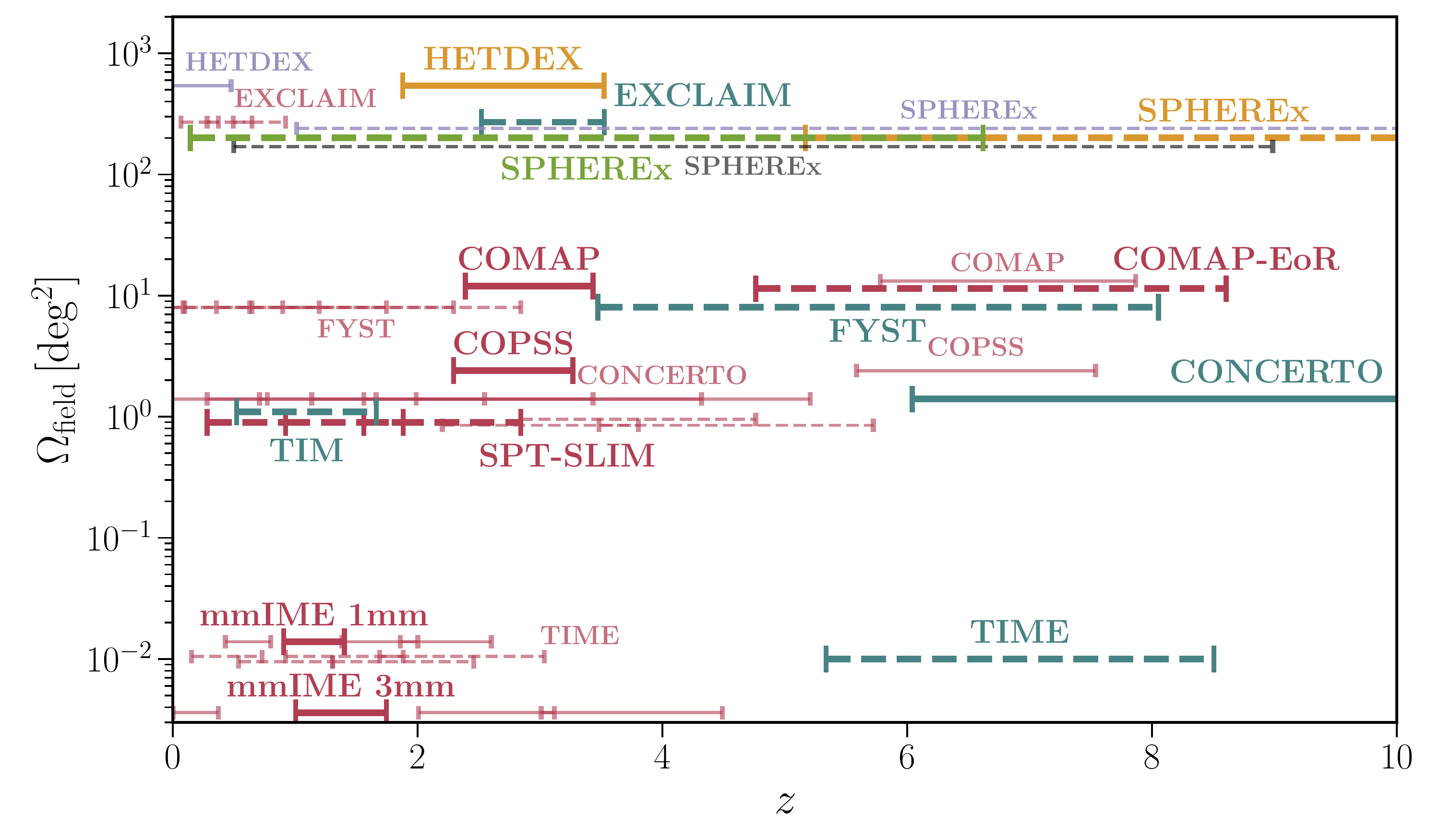}
\includegraphics[width=0.92\textwidth]{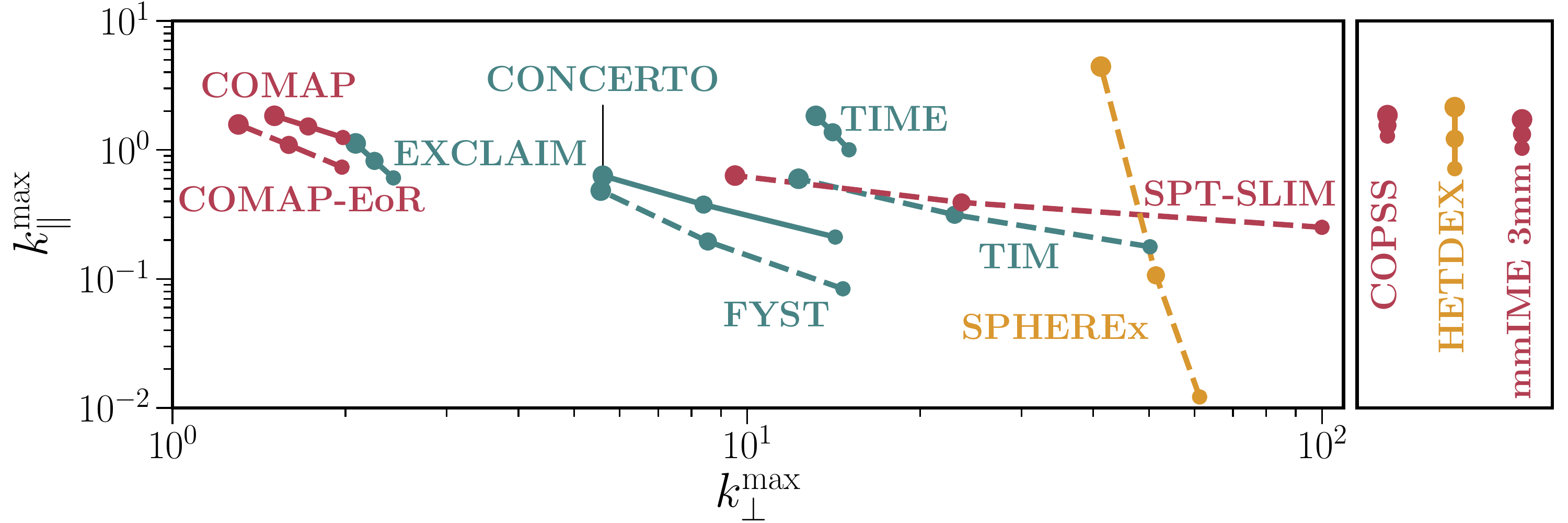}
\caption{Current experimental landscape of LIM, including a representative list of ongoing and funded LIM surveys, denoting the targeted (thick) and interloper (thin, semitransparent) lines with different colors. Top: redshift range and total sky coverage (summing different observed patches when required) probed by each experiment. In some cases, the nominal sky coverage is slightly modified to ease the reading of the figure.
Bottom: maximal line-of-sight and transverse wavenumbers accessible by each experiment with their main target emission lines, given the spectral and angular resolutions. We show values for the minimum, mean and maximum redshifts (marked by increasing size of the marker). The separate bottom panel contains experiments with significantly higher angular resolution. 
} 
\label{fig:EXP}
\end{centering}
\end{figure}

The vast range of targeted wavelengths necessitates the use of different LIM instruments. Hence, instrumental and observational challenges and sources of contamination are not shared by all experiments, which renders the joint scope of LIM observations cleaner. Over the coming decade, more than a dozen ground-based, balloon-borne and space satellite experiments are expected to deliver line-intensity maps spanning the full history of the Universe from the present day to the EoR. As shown in Fig.~\ref{fig:EXP}, LIM surveys will gradually cover larger sky areas, as the field transitions from the current pathfinder era of first detections, to an  early science era where they can be used to augment other measurements---particularly exploiting cross-correlation opportunities with other observables---and make advances in a plethora of areas in astrophysics and cosmology. Representatives of these eras include ongoing and funded experiments such as: COPSS~\cite{Keating:2015qva,Keating:2016pka}, mmIME~\cite{Keating:2020wlx}, COMAP~\cite{Cleary:2021dsp,COMAP:2021nrp}, FYST~\cite{CCAT-Prime:2021lly} and SPT-SLIM~\cite{Karkare:2021ryi}, targeting CO; CONCERTO~\cite{2020A&A...642A..60C}, TIME~\cite{2014SPIE.9153E..1WC,Sun:2020mco}, FYST and the balloon-borne EXCLAIM~\cite{2021JATIS...7d4004S} and TIM~\cite{2020arXiv200914340V}, targeting [CII]; HETDEX~\cite{Hill:2008mv,Gebhardt:2021vfo}, which will map Ly$\alpha$; and the SPHEREx satellite~\cite{2014arXiv1412.4872D}, which will measure [OII], [OIII], H$\alpha$ and Ly$\alpha$. Figure~\ref{fig:EXP} also shows the smallest scales accessible by this group of experiments, including their redshift dependence.

Eventually, a third generation of LIM experiments will map huge volumes by either covering large sky fractions, like one of the proposed surveys for AtLAST~\cite{2020SPIE11445E..2FK},  using high-sensitivity wide-bandwidth instruments (e.g.\ CDIM~\cite{Cooray:2016hro,2019BAAS...51g..23C}, COMAP-ERA~\cite{COMAP:2021nrp}), or both (see  ESA Voyage-2050 proposal~\cite{Silva:2019hsh, Delabrouille:2019thj}). Such flagship missions will truly unravel the potential of LIM~\cite{Kovetz:2019uss}. 

We hope these optimistic arguments about the prospects of LIM provide ample motivation for the reader to indulge further in this review as it goes into more detail. 
However, it is important to bear in mind that LIM also presents a different set of challenges compared to other observables, including limitations due to thermal detector noise, contamination from continuum emission or interloper lines (spectral lines redshifted to the observed frequencies from other cosmological volumes than that of the target lines), and degeneracies between astrophysics and cosmology. These various traits and the methods proposed to address them will be covered in depth in the following sections of this review. 

The outline of the review is as follows. We present an introduction to the primary emission lines targeted by LIM experiments in Sec.~\ref{sec:lines}, and discuss different strategies to model their intensities in Sec.~\ref{sec:modeling}. After fully introducing LIM, we survey the prospects of this technique to improve our understanding of astrophysics and cosmology in Sec.~\ref{sec:prospects}. We then lay out the formalism for describing the statistical properties of line-intensity maps as well as the potential of cross-correlations with external observables in Secs.~\ref{sec:formalism} and~\ref{sec:crosscorrelation}, respectively. We close the review with brief concluding remarks in Sec.~\ref{sec:conclusions}.

\section{Lines}
\label{sec:lines}

A variety of galactic emission lines ranging from the microwave to the ultraviolet (UV) bands can be used to probe different phases of the IGM and ISM, and to study the various astrophysical processes shaping the host galaxies~\cite{2019ARA&A..57..511K}. Line-intensity maps are connected to the emission from stars in different ways. Stellar emission ionizes and heats the ISM, which absorbs part of the UV radiation and re-emits it in the infrared~\cite{2012ARA&A..50..531K, Madau:2014bja}. Thus, as we  discuss in Sec.~\ref{sec:modeling}, the relation between the UV and infrared luminosities is empirically related to the stellar mass or the UV-continuum spectral index~\cite{Heinis:2013dsa, 2016ApJ...833...72B, 2020ApJ...902..112B}. The stellar emission and infrared re-emission determine the ionization and photo-dissociation structure of the galaxy, triggering (directly or through  cascades of events) the luminosity of the lines that LIM experiments target, as we detail below. 

\begin{figure}[h!]
 \begin{centering}
\includegraphics[width=\textwidth]{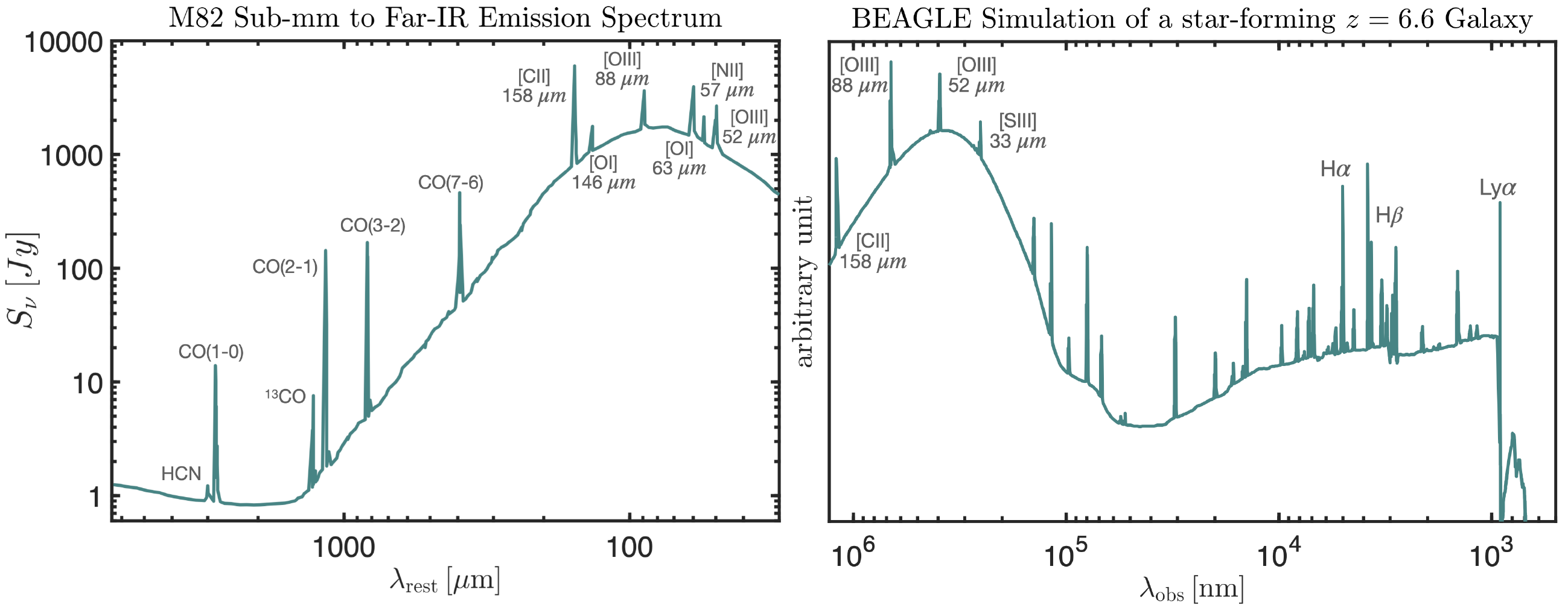}
\caption{{\it Left:}  M82 emission spectrum, at rest-frame wavelengths. 
Adapted from Ref.~\cite{Wilson:2006cn}. {\it Right:} A model spectral energy distribution from the BEAGLE simulation~\cite{2016MNRAS.462.1415C}, at the observed wavelengths, of a star-forming galaxy at redshift $z\!=\!6.6$.  Adapted from Ref.~\cite{2018ApJ...859...84H}. We highlight some of the main lines targeted by LIM experiments, from CO(1-0) to Ly$\alpha$. 
}
\label{fig:submmlines}
\vspace{-0.2in}
\end{centering}
\end{figure}

In this Section, we briefly introduce the main target lines for LIM 
following an increasing order in frequency, starting in the sub-millimeter range. In Fig.~\ref{fig:submmlines} we show examples of galactic spectra from the sub-millimeter to the near-UV.

\subsection{Carbon Monoxide (CO)}
\label{sec:CO}

CO is the most common molecule in the Universe besides diatomic molecular hydrogen (H$_2$) 
and is the most widely-used tracer of 
molecular gas~\cite{1987ApJ...319..730S, Dame:2000sp, Walter:2003zh,2011MNRAS.412.1913I, 2011MNRAS.415...32S,Carilli:2013qm}.
Its rotational line emissions, at a ladder of frequencies 
$\nu_{J\rightarrow J-1}=J\times115.27\, {\rm GHz}$
(or wavelengths $\lambda=2.6/J\,{\rm mm}$)
for $(J \rightarrow J-1)$ transitions, 
are among the brightest in galactic spectra and can be efficiently observed by 
terrestrial telescopes targeting the sub-mm wavelength range (including the higher rotational lines that originate from high-redshift sources). 

In principle, the CO(1-0) luminosity from virialized molecular clouds is linearly related to the cloud H$_2$ mass~\cite{Narayanan:2011zm,2012ARA&A..50..531K,Bolatto:2013ks,2015ARA&A..53..583H} (provided the volume covered by molecular clouds is small enough so that the emission from one cloud is not absorbed by another~\cite{1986ApJ...309..326D, Bolatto:2013ks}), which can be used to estimate the amount of stellar mass and the star-formation rate in galaxies~\cite{Tacconi:2020wdz}. However, the CO emission is very sensitive to the environment~\cite{2012A&A...541A..58L}, depending on various factors such as metallicity~\cite{Genzel:2011cw,Popping:2013ixa, 2022arXiv220406937S}, gas temperature and density~\cite{Krumholz:2011mq}, the existence of a starburst phase~\cite{Downes:1998vm,Narayanan:2005xj}, and the destruction of CO by cosmic rays~\cite{Papadopoulos:2011kb,2015ApJ...803...37B}.

All this makes the interpretation of CO LIM observations as a proxy for cosmic molecular abundance challenging~\cite{Breysse:2021ecm}, although there are ways to tighten this relationship. For example, 
luminosity ratios between different CO lines provide useful constraints on the physical conditions in the gas~\cite{Solomon:2005xc}. 
There is also potential to scrutinize the CO to H$_2$ relation directly using LIM of hydrogen deuteride, which may be observable during reionization with future instruments~\cite{Breysse:2021utr}. Another interesting target is the $^{13}$CO isotopologue at $\nu_{\rm rest}=110\,{\rm GHz}$~\cite{Breysse:2016opl};  cross-correlating $^{13}$CO and $^{12}$CO from the same sources  provides an estimate of the gas density, as the former saturates at a much higher column density. Finally, due to similar critical densities, fine-structure lines of CI are highly correlated with CO lines independently of the environment, providing an alternative tracer of the molecular gas~\cite{2015A&A...578A..95I, Jiao:2017knu, 2018ApJ...869...27V, 2019A&A...624A..23N}, which can also be targeted by LIM~\cite{Sun:2020mco, Chung:2022lpr,Bethermin:2022lmd} and used to break degeneracies related with gas temperatures, excitation states and column densities. 

As it has the lowest frequency among bright emission lines and is quite far from the HI line, observations of the CO(1-0) line are not prone to contamination from foreground interloper lines. The main culprit, HCN~\cite{Breysse:2015baa}, is quite weak in comparison~\cite{Chung:2017uot} (see Fig.~\ref{fig:submmlines}). However, higher CO rotational lines can be mixed with lower ones, and many of them are strong interlopers for [CII] measurements at high-redshift.

\subsection{Ionized Carbon [CII]}
\label{sec:CII}

Atomic and ionic fine-structure lines in the infrared are important drivers of the cooling process 
of interstellar gas~\cite{Carilli:2013qm}. The [CII] $158\,{\mu \rm{m}}$ fine-structure line, mostly emitted from dense photo-dissociation regions in the outer layer of molecular clouds~\cite{Stacey:2010ps}, is the brightest among them~\cite{Tielens:1985st, 1991ApJ...373..423S, Wolfire:2022dbc} (see Fig.~\ref{fig:submmlines}). There are studies  indicating that the [CII] emission can also trace molecular gas~\cite{2018MNRAS.481.1976Z}, even outperforming CO(1-0) for high-redshift low-metallicity galaxies~\cite{2022arXiv220305316V}, and recent 
works suggest  that a small but non-negligible fraction of the [CII] radiation may be emitted from ionized gas phases~\cite{2015A&A...575A..17H, 2017ApJ...845...96C, 2019A&A...626A..23C}. 

Assuming that galactic dust converts the vast majority of the UV and optical radiation it absorbs into infrared luminosity, and that only a remainder fraction is applied to photo-electric heating, the [CII] luminosity is proportional to the heating rate and to the fraction of atomic gas.\footnote{This simplified description ignores the role of fainter cooling lines~\cite{Tielens:1985st, 2002ApJ...578..885Y}, the dependence of the photo-electric efficiency of dust grains on their charge~\cite{1994ApJ...427..822B}, and the saturation of the [CII] line at high temperatures and radiation intensities~\cite{2016MNRAS.463.2085M, 2019ApJ...876..112R}.} 
Thus, [CII] provides a very natural target for LIM experiments to trace the star-formation history~\cite{Suginohara:1998ti,DeLooze:2011uw,Herrera-Camus:2014qba}, and due to its brightness, it is especially targeted at high redshifts. However, the tight relationship between the [CII] line and infrared luminosities yields a much larger scatter at high redshifts compared to  what is observed at lower redshifts. Possible explanations~\cite{2019MNRAS.489....1F,2022arXiv220508905B} include a large population of galaxies undergoing starbursts~\cite{2015ApJ...813...36V}, low metallicities~\cite{2015ApJ...813...36V,2017ApJ...846..105O,2018A&A...609A.130L}, and intense radiation fields in compact galaxies that photo-evaporate molecular clouds~\cite{Gorti:2002xqa,2017MNRAS.471.4476D,2019MNRAS.487.3377D},  regulating the [CII] luminosity~\cite{2017MNRAS.467.1300V}.\footnote{This has also been invoked to explain the [CII] deficit in some local galaxies~\cite{2017ApJ...846...32D,2017MNRAS.467...50N,2018ApJ...861...95H}.}  The effects of these processes are degenerate in the [CII] luminosity, but cross-correlating [CII] maps with observations of other far-infrared, CO, optical or UV lines, or with the cosmic infrared background (CIB), will help break the degeneracies.

The frequency of the [CII] line lies just above the ladder of CO lines, hence it suffers from their contamination as foreground line-interlopers~\cite{Breysse:2015baa}. Other atomic fine-structure lines (such as ionized oxygen and nitrogen or neutral carbon) also contaminate the [CII] line-intensity maps, but typically less severely.

\vspace{-0.075in}

\subsection{Other Atomic Fine-Structure lines}
\label{sec:other_finestructure}

Besides the carbon lines, other far–infrared fine-structure lines that can be used to probe ISM physics include  silicon [SIII] $18\,{\rm \mu m}$ and $33\,{\rm \mu m}$, oxygen [OI] $63\,{\rm \mu m}$, [OIII] $52\,{\rm \mu m}$ and $88\,{\rm \mu m}$, and nitrogen [NII] $122\,{\rm \mu m}$ and $205\,{\rm \mu m}$.
These lines and the ratios between them and other lines such as [CII] provide additional means to  measure the electron density, excitation temperatures, gas pressure, metallicity, ionization parameter, the hardness of the ionizing radiation and the properties of both the neutral and ionized gas phases~\cite{Serra:2016jzs,2019ARA&A..57..511K,2019ApJ...887..142S,2020MNRAS.499.3417Y,2021MNRAS.504..723Y,Padmanabhan:2021tjr,Padilla:2022asq}.

For example, although [OIII] requires hard ionizing radiation and is typically weaker than [CII], in certain environments such as AGN and early galaxies it can in fact be brighter~\cite{Carilli:2013qm} (see Fig.~\ref{fig:submmlines}). 
High  [OIII]/[CII] ratios can hint at a [CII] deficit~\cite{Laporte:2019rbp}, but can also result from underestimation of [CII] emission in targeted observations as its emission tends to be more extended than [OIII]~\cite{Carniani:2020ldd}. Another example involves [NII], a tracer of regions of ionized hydrogen; the [NII] line luminosities depend on the electron density, the ionized-gas temperature,  and the nitrogen-to-hydrogen abundance ratios~\cite{2015ApJ...814..133G,2016ApJ...826..175H}.

\vspace{-0.075in}

\subsection{Optical and ultraviolet hydrogen lines}
\label{sec:UV}

Several key hydrogen lines such as Ly$\alpha$, H$\alpha$ and H$\beta$, emitted in the UV and optical and redshifted to wavelengths down to the infrared (see Fig.~\ref{fig:submmlines}), provide another set of important targets of multiple  LIM experiments.  

Hydrogen recombinations following ionization by young stars and AGN, as well as collisional excitations from shock heating and cold accretion, can result in Ly$\alpha$ line emission, the most energetic line emission  from star-forming galaxies~\cite{Ouchi:2020zce}. As they traverse the ISM and IGM, Ly$\alpha$ photons get repeatedly absorbed and re-emitted by neutral hydrogen. This multiple scattering disperses their directions and frequencies~\cite{Santos:2003pc} and increases the probability that they get absorbed by galactic dust. Thus, galaxy metallicity and dust content both have a direct impact on the observed Ly$\alpha$ emission~\cite{Ouchi:2020zce}. In general, the escape fraction of Ly$\alpha$ photons decreases with higher star-formation rate (which is associated with higher dust abundance~\cite{Santini:2013yfa}), and increases, for high-redshift galaxies, with their redshift. As higher redshifts are probed, Ly$\alpha$ emission, especially through its 
escape fraction, provides an effective tracer of the neutral gas abundance. This is especially true for the dimmer emission from neutral circumgalactic and intergalactic media, which eventually become the tail of the EoR, and are more accessible with LIM~\cite{Steidel:2011ey,Zheng:2010tw}. Hence,  intensity maps of localized and extended Ly$\alpha$ emissions correlate with star-forming lines at the center of the ionizing sources, e.g.\ CO and [CII]~\cite{Beane:2018pmx,Moriwaki:2019dbg}, as well as with the HI line from regions surrounding the bubbles~\cite{Sobacchi:2016mhx}.   

The lower frequency hydrogen lines H$\alpha$ and H$\beta$ are emitted from the same sources as Ly$\alpha$ as a result of the  ensuing radiative cascade of  Ly$\alpha$ excitations.  They are also susceptible to dust extinction, but much less than Ly$\alpha$~\cite{2017arXiv171109902S}. Roughly three times brighter than H$\beta$~\cite{2006agna.book.....O},  H$\alpha$ provides a fairly direct probe of star formation~\cite{Kennicutt:1998zb, Ly:2006hx,Saito:2020qxq} and can also be used in cross-correlation.  For example, using LIM to measure the ratio between HeII (1640 ${\rm \AA}$) and H$\alpha$ towards cosmic dawn redshifts, $z\lesssim20$, can constrain the initial mass function of Pop III stars~\cite{Visbal:2015sca,Parsons:2021qyw}, as these produce more HeII ionizing photons than metal enriched stars~\cite{Schaerer:2001jc}.

Although the UV hydrogen lines have a series of interlopers~\cite{Fonseca:2016qqw, Gong:2020lim} (see Fig.~\ref{fig:submmlines}), the resolution required for these high-frequency observations typically allows separating between the target and interloper lines~\cite{Breysse:2015baa,Pozzetti:2016cch,2017arXiv171109902S}. Some of these metal interlopers, such as [OII] 373.7 nm, [OIII] 495.9 nm and 500.7 nm, and [NII] 655.0 nm, provide faithful tracers of the star formation rate at low redshifts~\cite{Kennicutt:1998zb, Ly:2006hx,Villa-Velez:2021ojy}.  Moreover, the [OII] doublet in particular can yield precise redshifts for galaxies observed with low integration times as is critical for the emission-line galaxy samples of eBOSS~\cite{Raichoor:2017nuz} and DESI~\cite{Raichoor:2020jcl}.

\vspace{-0.175in}

\subsection{Preliminary detections}
\label{sec:detections}

\vspace{-0.05in}

Risking a subsection that will quickly become obsolete, we briefly summarize the preliminary detections of LIM to date, to provide context for what follows.

CO(1-0) at redshifts $1\!<\!z\!<\!5$ has been the target line for several pathfinder 
LIM experiments. 
The first $\sim\!2\sigma$ detection of its shot-noise power  at redshift $z\!\sim\!3$ was achieved by the COPSS
survey using data from the Sunyaev-Zel'dovich Array~\cite{Keating:2015qva,Keating:2016pka}. This
was followed by a stronger $\sim\!4\sigma$ detection by  mmIME~\cite{Keating:2020wlx}, using data from ALMA~\cite{2019ApJ...882..138D, 2019ApJ...882..139G} and  ACA~\cite{Scoville:2006vq}. Recent work~\cite{Keenan:2021uue} showed the promise of cross-correlating CO(1-0) with galaxy surveys (and obtained preliminary upper limits at $z\sim 3$).

The first [CII] LIM measurement, at $2\sigma$ confidence-level, was obtained via cross correlation between  maps from the Planck High Frequency Instrument and high-redshift catalogs of quasars and luminous red galaxies (LRGs)~\cite{Pullen:2017ogs}, with an improved methodology yielding a $\sim 4\sigma$ detection~\cite{Yang:2019eoj}. The detected excess above the CIB, stellar radiation reprocessed as infrared continuum emission by the dust~\cite{Hauser:2001xs, Kashlinsky:2004jt,Wu:2016vpb},  
appears to be consistent with collisional excitation models of [CII] emission, where the intensity is proportional to the collisional rate, which depends 
on the gas density and temperature.

The Ly$\alpha$ LIM signal has been searched for using stacking~\cite{Niemeyer:2022vrt} and cross correlation between source redshift catalogs and maps expected to include residual Ly$\alpha$ emission from the same sources. First attempts to cross correlate BOSS quasar catalogs with BOSS LRG spectra proved challenging~\cite{BOSS:2015ids,Croft:2018rwv,Renard:2020mfg}. More recently, a $\sim\!3\sigma$ detection was reported using stacking of Subaru/HSC narrow-band images at $z\sim 6$ around resolved bright Ly$\alpha$ emitters~\cite{Kakuma:2019afo}.

\section{Modeling}
\label{sec:modeling}

LIM experiments measure the specific intensity $I$ per unit of observed frequency $\nu_{\rm obs}$,\footnote{The specific intensity is sometimes denoted with $I_\nu$ to distinguish it from the integrated intensity $I_\nu{\rm d}\nu$. To simplify the notation, 
we do not follow this convention and use $I$ to refer to specific intensities throughout this review.} which can be derived from the line-luminosity density $\rho_{\rm L}$ per comoving volume. We can transform $\rho_{\rm L}$  into flux volume density dividing by $4\pi D_L^2$, which is in turn converted to specific intensity by transforming the comoving volume element to solid angle and observed frequency elements as ${\rm d}A/{\rm d}\Omega=D_M^2$ and ${\rm d}\chi/{\rm d}\nu_{\rm obs}=c(1+z)/H\nu_{\rm obs}$, respectively. In the explanation above, $D_L$ is the luminosity distance, $D_M$ is the comoving angular diameter distance, $\chi$ is the comoving radial distance, $c$ is the speed of light and $H(z)$ is the Hubble expansion rate. However, experiments covering frequencies below some tens of GHz usually employ the brightness temperature $T=c^2I/(2k_B\nu_{\rm obs}^2)$ using the Rayleigh-Jeans relation. Therefore, we can use
\begin{equation}
     I(z) = \frac{c}{4\pi\nu H(z)}\rho_{\rm L}(z)\,,\qquad {\rm or} \qquad T(z) = \frac{c^3(1+z)^2}{8\pi k_{\rm B}\nu^3H(z)}\rho_{\rm L}(z)\,,\quad
\label{eq:def_IandT}
\end{equation}
where $\nu$ is the line rest-frame frequency, and $k_{\rm B}$ is the Boltzmann constant.

Contrary to the HI emission before reionization, the spectral lines we focus on in this review originate in galaxies and the IGM within dark matter halos, given their relation to emission from young, bright stars. Therefore, it is fair to identify the line sources with galaxies, although in some cases, especially for the Ly$\alpha$ line, radiative transfer extends the emission profile far beyond the size of the halo~\cite{2021ApJ...916...22K, 2022ApJ...929...90L, 2021arXiv210809288K, Niemeyer:2022vrt}. $\rho_{\rm L}$ can be obtained using either a luminosity function or a direct relation $L(M,z)$ between the halo mass and the line luminosity at each redshift. For instance, the mean luminosity density can be obtained as
\begin{equation}
    \langle \rho_{\rm L}\rangle(z) = \int {\rm d}ML(M,z)\frac{{\rm d}n}{{\rm d}M}(M,z) = \int{\rm d}L L \frac{{\rm d}n}{{\rm d}L}(z)\,,
\label{eq:mean_rhoL}
\end{equation}
where ${\rm d}n/{\rm d}M$ and ${\rm d}n/{\rm d}L$ are the halo-mass and line-luminosity functions. From expressions like this and through the connection to the halo distribution, it is possible to derive summary statistics to describe line-intensity maps and extract astrophysical and cosmological information. In Sec.~\ref{sec:formalism} we will focus on one- and two-point statistics (the voxel-intensity distribution (VID) and the power spectrum, respectively). We will use $L(M)$ in what follows unless otherwise stated, since it allows for more direct physical modeling of the line intensity, although all of our expressions can be easily adapted to the luminosity function. To model the line intensity, we need to relate it with the galaxy properties (this can be compressed in the $L(M)$ relation thereafter).

There are two main approaches to estimating 
the line intensity in connection to the galaxy properties. One possibility relies on empirical  scaling relations fit to observations. Another  involves 
theory-motivated relations, that may range from analytic derivations to semi-analytic models to fully-fledged hydrodynamic simulations. For the sake of usability, empirical relations can also be calibrated on the results of these simulations rather than observations. Each approach has its own benefits and flaws, and a robust understanding of the astrophysical dependence of LIM will likely require a combination of both. 

\begin{figure}[h!]
 \begin{centering}
\includegraphics[width=\textwidth]{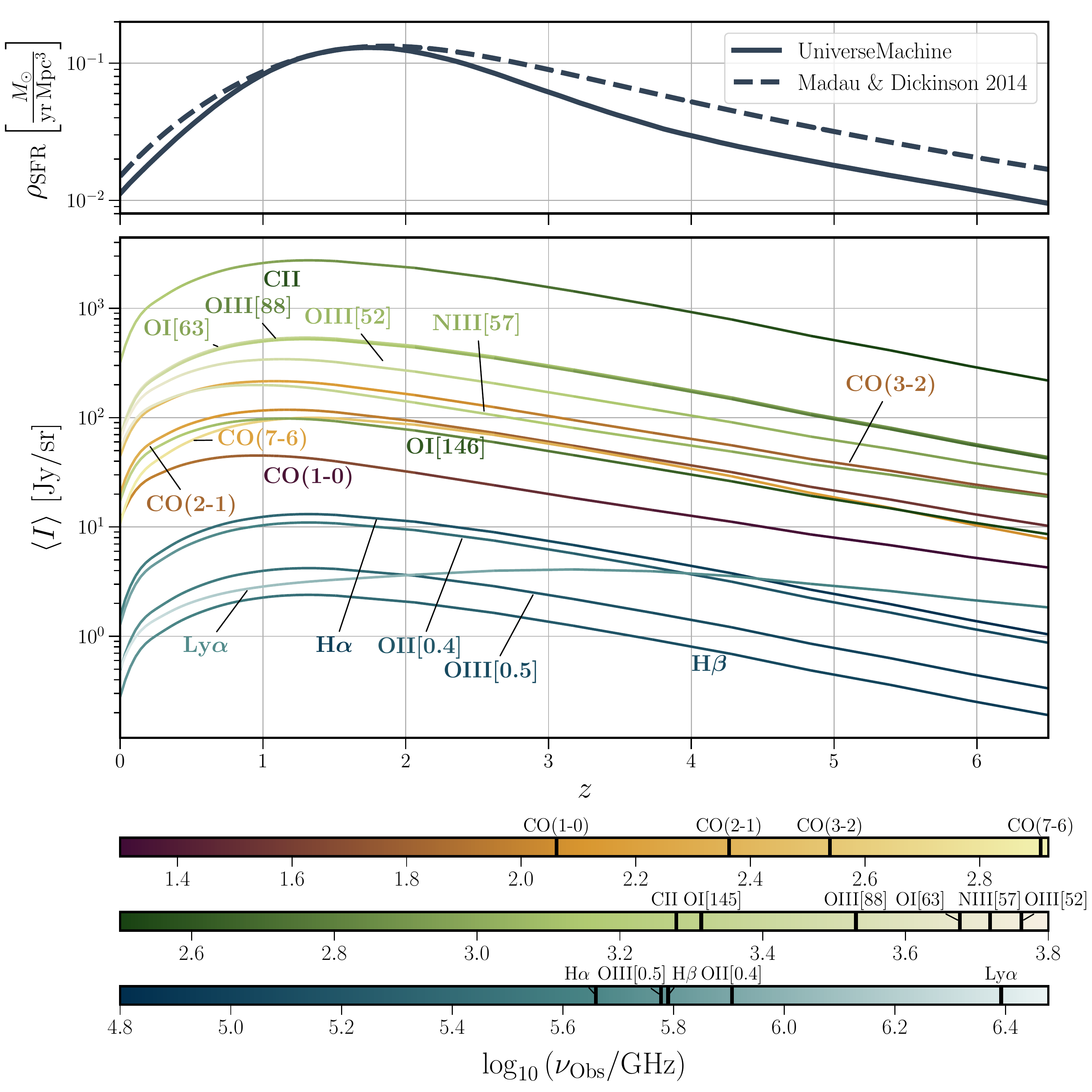}
\caption{\textit{Bottom:} Microwave, infrared, optical and UV spectral lines (rest-frame frequencies marked with vertical bars in the colorbar) mean intensity as function of redshift, obtained using scaling relations from the infrared luminosity and star formation rates~\cite{Kennicutt:1998zb,2016ApJ...829...93K,Spinoglio:2011ug,DeLooze:2014dta,2017ApJ...835..273G, COMAP:2018svn}. We use star-formation rates and quenching fractions from Refs.~\cite{Behroozi:2012iw,Behroozi:2019kql} assuming a mean-preserving logarithmic scatter of $0.3\,{\rm  dex}$ in the star-formation rate to halo mass relation. \textit{Top:} Star-formation rate density~\cite{Behroozi:2019kql,Madau:2014bja}. Line-intensity ratios evolve with redshift; current observational constraints apply at specific redshifts and are uncertain by up to one order of magnitude.}
\label{fig:kernels}
\end{centering}
\end{figure}

In Fig.~\ref{fig:kernels} we show the results of using empirical scaling relations to model the  intensities as a function of redshift for the LIM target lines discussed above.

\subsection{Empirical Scaling relations}
\label{sec:modeling_observations}

Scaling relations are  most useful when they connect line luminosities to quantities that are relatively easy to observe, such as the infrared luminosity, or to general galaxy properties that can be inferred from observations, like the star-formation rate. For instance, since optical hydrogen and oxygen lines are good star-formation tracers, their luminosities are in tight linear relations with the star-formation rate, as shown in observations~\cite{Kennicutt:1998zb, Ly:2006hx, 2011ApJ...737...67M,Villa-Velez:2021ojy}, and based on stellar evolutionary tracks, ionizing emissivity and recombination rates.

Meanwhile, the emission of the [CII] line correlates with both the far-infrared emission from dust~\cite{1985ApJ...291..755C, 1991ApJ...381..200W} and with star formation~\cite{1991ApJ...373..423S,DeLooze:2014dta,2017ApJ...834...36H}. Low-redshift~\cite{DeLooze:2014dta,Herrera-Camus:2014qba} and high-redshift~\cite{2015Natur.522..455C,Carniani:2020ldd,2020A&A...643A...3S} observational studies obtained a nearly linear correlation between [CII] luminosity and the star-formation rate, without redshift dependence, while numerical simulations~\cite{2015ApJ...813...36V,2017ApJ...846..105O, 2020MNRAS.492.2818L, Pallottini:2022inw} and semi-analytic models~\cite{2018A&A...609A.130L, 2019MNRAS.482.4906P} predict a weak dependence on redshift.  Observational correlations between the [CII] luminosity and star-formation rates in the local Universe~\cite{DeLooze:2014dta} are found to still apply at high redshifts~\cite{2015Natur.522..455C,Chung:2018szp,ALPINETeam:2019arv,2020ApJS..247...61F}. This is surprising, 
since massive high-redshift galaxies correspond to a much higher [CII] surface brightness regime than the populations observed at low-redshift~\cite{2016ApJ...833...71A,2019MNRAS.489....1F,Pizzati:2020bqq}, and may change with more observations.

Fine-structure lines, as dust-cooling emission lines, are correlated to the infrared luminosity~\cite{Spinoglio:2011ug, 2017ApJ...846...32D}. The situation is very similar for the ladder of CO rotational lines, with the subtle difference of observational results being reported in terms of the $L'$ pseudo luminosity, expressed in K km/s pc$^2$ units. There is rich literature on relations for the ground transition~\cite{Daddi:2010nn, Carilli:2013qm, 2014ApJ...794..142G,Sargent:2013sxa, 2015A&A...577A..50D}. The intensity of the other CO rotational transitions is determined by the spectral line energy distribution, which shows large variety between different studies~\cite{2015ApJ...801...72R, 2016ApJ...829...93K, 2017A&A...608A.144Y,2018A&A...620A..61C,2020A&A...641A.155V,2020ApJ...902..109B, 2020ApJ...889..162L}. The spectral-line energy distribution depends largely on the galaxy type~\cite{2016ApJ...829...93K}, with the CO to infrared-luminosity relations typically lower in starburst galaxies compared to main-sequence galaxies~\cite{Aalto:1995au,Genzel:2010na,Daddi:2010nn,2015ApJ...810L..14L},  mainly due to differences in star-formation to molecular mass ratios.

As an example, let us briefly describe the procedure to obtain $L(M)$ for the CO lines.\footnote{A detailed discussion of a popular model for the CO line can be found in Ref.~\cite{Li:2015gqa}.} The first step is to assign a star-formation rate $\rm{SFR}(M,z)$ to any halo of mass $M$ at redshift $z$. The mean $\rm{SFR}(M,z)$ is often parameterized with a double power law~\cite{Silva:2014ira, Fonseca:2016qqw,2017ApJ...835..273G}, or using more involved fitting functions~\cite{Behroozi:2019kql}. 
As empirical relations are not ubiquitous and there is significant variability in the populations considered, 
it is quite common to add a characteristic log-normal scatter\footnote{A log-normal scatter may not describe the whole population distribution accurately. For instance, star-forming and quenched galaxy populations can each introduce their own scatter, typically resulting in a bimodal distribution~\cite{Behroozi:2019kql}.}. 
A typical value is $\sigma_{\rm SFR}\sim\! 0.3\, {\rm dex}$~\cite{Behroozi:2012iw}. 
In the second step, we assume that the infrared luminosity and the star-formation rate are correlated, and adopt an ansatz $L_{\rm IR} = 1.72\times 10^{-10}{\rm SFR}/M_{\odot} {\rm yr}^{-1} L^{-1}_{\odot}$~\cite{Kennicutt:1998zb}. Finally, we use a power-law relation between infrared and CO luminosities 
\begin{equation}
    \log L_{\rm IR} = \alpha_J \log L'_{\rm CO(J\rightarrow J-1)} + \beta_J\,,
\label{eq:IR-to-CO}
\end{equation}
where $\alpha_J$ and $\beta_J$ depend on the transition and the data used to calibrate the correlation, and $L_{\rm CO(J\rightarrow J-1)}/L_\odot = 4.9\times 10^{-5}J^3(L'_{\rm CO(J\rightarrow J-1)}/\rm K\ km\ s^{-1}\ pc^2)$. We then add another log-normal scatter $\sigma_{\rm L}\sim\! 0.3\,{\rm dex}$. To summarize, we assign
\begin{equation}
    L_{\rm CO(J\rightarrow J-1)}\Bigg{(}L_{\rm IR}\Big{(}{\rm SFR}(M,z),\sigma_{\rm SFR}\Big{)},\alpha_J,\beta_J,\sigma_{\rm L}\Bigg{)}
    \label{eq:LCO_empirical}
\end{equation}
to any halo of mass $M$ at redshift $z$.
We note that this relation can be made more accurate if the stellar mass $M_*$ in the halo can be estimated. In this case we can use the observed infrared-to-ultraviolet excess (IRX$\equiv L_{\rm IR}/L_{\rm UV}$) to account for the stellar light that is not reprocessed into infrared photons
\begin{equation}
    \text{SFR} = K_{\rm UV} L_{\rm UV} + K_{\rm IR} L_{\rm IR}\,,\qquad\qquad \text{IRX} = \left(\frac{M_*}{M_s}\right)^{x}\,,
\end{equation}
where $K_{\rm UV} = 2.5\times 10^{-10}\ M_{\odot} {\rm yr}^{-1} L^{-1}_{\odot}$ and $K_{\rm IR} = 1.73\times 10^{-10}\ M_{\odot} {\rm yr}^{-1} L^{-1}_{\odot}$ (following Ref.~\cite{Madau:2014bja}),\footnote{These values correspond to a Salpeter initial mass function and must be multiplied by 0.63 to convert them to the Chabrier initial mass function.} and  $x=0.97^{+0.17}_{-0.17}$, $\log_{10}(M_s/M_{\odot}) = 9.15^{+0.18}_{-0.16}$, and a log-normal scatter of $\sigma_{\rm IRX} = 0.2$ dex~\cite{2020ApJ...902..112B} is included.\footnote{Fitting formulae relating IRX to the spectral index of the UV-continuum emission are also available.} This improved relation downweights the contribution from halos with low star-formation rate~\cite{Wu:2016vpb}.

Ly$\alpha$ emission involves more complicated radiative transfer, including dust absorption without photo-ionization, recombination line emission absorbed by dust, and ionizing photons escaping the galaxy without any ionization events or being absorbed by HI regions without triggering recombination line emission. For the purposes of Ly$\alpha$ intensity mapping, these processes can be merged and modeled as an escape fraction, which depends significantly on the environment and is all but unconstrained observationally. However, current measurements show two general trends~\cite{COMAP:2018svn}: the escape fraction increases monotonically with redshift, and it decreases with higher star-formation rate. 
One can then use a general function depending on redshift and star-formation rate satisfying these limiting trends~\cite{COMAP:2018svn}, or other parameterizations based on constraints to the escape fraction and other contributions to the Ly$\alpha$ intensity~\cite{Fonseca:2016qqw, Silva:2012mtb}.

Using empirical relations provides a fast method to predict the line intensities, taking into consideration observational constraints and without relying on theoretical priors, hence involving a more agnostic prediction accounting for potentially unknown astrophysics. 
Empirical approaches can be very powerful in resolving occasional discrepancies between theory-based estimates.  
On the other hand, the observations used to calibrate these scaling relations span a very specific redshift interval, which either limits their application to LIM analyses or forces to extrapolate the results to other redshifts. The latter option  may result in very inaccurate estimations. One example of this is the ground transition of CO~\cite{Breysse:2014uia}, as highlighted in Ref.~\cite{COMAP:2021rny}. Using low-redshift data~\cite{Carilli:2013qm, 2016ApJ...829...93K} one finds best-fit values around $\alpha_1\approx 0.1\beta_1+1.19$~\cite{Li:2015gqa}   for the relationship between the parameters in Eq.~\eqref{eq:IR-to-CO}, while using high-redshift observations, 
such as the COLDz luminosity functions at $z\sim 2.4$~\cite{Riechers:2018zjg}, yields $\alpha_1=0.67$ and $\beta_1=4.90$~\cite{LC_paper}.

\subsection{Theory-motivated approaches}
\label{sec:modeling_theory}

There are three main types of theory-motivated approaches to model line intensities, which can be distinguished by their numerical complexity: analytic models, semi-analytic models and 
hydrodynamic simulations. 

Analytic models aim to relate line luminosities to the ISM properties,  
which requires modeling the phases of the galaxy and the relative abundances of each of its main components. There are many ways to achieve this goal, but all of them start with modeling the stellar radiation. Some options~\cite{2019ApJ...887..142S} use CIB models~\cite{Shang:2011mh} and relations between star-formation rate and infrared luminosities~\cite{Kennicutt:1998zb}. The gas-to-dust mass ratios can be obtained from the dust model using observation-based calibrations of the dust-grain emission~\cite{2014A&A...566A..55P}, in addition to the total hydrogen mass, which is distributed between neutral, ionized and molecular hydrogen~\cite{2020ApJ...902..111W}. Finally, the metallicity is obtained from the hydrogen and dust masses, assuming a constant dust-to-metal ratio (which can be obtained from hydrodynamic galaxy formation simulations~\cite{2019MNRAS.490.1425L}). Other options~\cite{2015JCAP...11..028M, 2019MNRAS.489....1F} model the ionization and photo-dissociation structure properties of a galaxy to determine its ionized and neutral regions as a function of the galaxy's stellar emission and ionization field. From this point, it is possible to estimate line luminosities in a general way that can be applied to both low and high redshift galaxies.

Analytic models are forced to make simplifying assumptions about the processes involved in the line emission for the sake simplicity. 
Concrete interstellar conditions can be taken into account with photo-ionization simulation codes~\cite{2017RMxAA..53..385F,Krumholz:2013qza, 2020MNRAS.494.1919L}, but in order to fully track galaxy properties, numerical hydrodynamic simulations are required.  These simulations 
evolve astrophysical properties self-consistently in each simulation cell, according to specific robust underlying physical baryonic models. 
However, the increased complexity sets an upper limit on the volume of the simulation as a tradeoff; if the simulation volume is too large, more of the baryonic physics implemented may rely on sub-grid models. The results of these simulations (see e.g.,~\cite{Hopkins:2013vha,McAlpine:2015tma,2015ApJ...813...36V,2017ApJ...846..105O,Nelson:2018uso,Dave:2019yyq,2020MNRAS.492.2818L,Pallottini:2022inw}) can be post-processed to consistently predict the intensity of several lines~\cite{2021arXiv211105354S}. Furthermore, the THESAN project~\cite{Kannan:2021xoz} has recently modeled several emission lines at reionization using a radiation-magneto-hydrodynamic simulation that self-consistently models hydrogen reionization and the  properties of the galaxies and active galactic nuclei~\cite{Kannan:2021ucy}. In practice this relies on several ad-hoc assumptions, such as a fixed ionization parameter, for instance.

Nonetheless, hydrodynamic simulations are computationally very expensive, which imposes significant limitations on their use and flexibility.  Semi-analytic models, in turn, offer a compromise between analytic approaches and hydrodynamic simulations~\cite{2015ARA&A..53...51S}. 
This approach dynamically evolves dark matter and baryons in a cosmological context, using motivated approximations relying on sub-grid physics to treat star formation and baryonic feedback (see e.g., Refs.~\cite{Somerville:1998bb, Lu:2013mxa, 2013MNRAS.431.3373H, Somerville:2008bx, 2015MNRAS.453.4337S, 2016MNRAS.462.3854L, Croton:2016etl}). The free parameters of these recipes are calibrated to match global observational quantities to observations or numerical simulations. Then, the results can be used to simulate multiple line luminosities for each galaxy~\cite{Lagos:2012sv, 2016MNRAS.461...93P, Dumitru:2018tgh, 2018A&A...609A.130L,2019MNRAS.482.4906P,  2020ApJ...905..102L}, and match it with a halo catalog within a lightcone from N-body simulations to build a mock LIM observation~\cite{2021ApJ...911..132Y}. 

It has been shown that hydrodynamic simulations and semi-analytic models are generally in good agreement~\cite{2015ARA&A..53...51S},\footnote{Although there are still significant discrepancies among the gas properties and star-formation efficiencies (see e.g.\ Refs.~\cite{2018MNRAS.474..492M,2012MNRAS.419.3200H}).} which motivates the use of semi-analytic models to generate simulated line-intensity maps. 
Finally, the results obtained from hydrodynamic simulations or semi-analytic models can be empirically summarized in scaling relations of the same form as the ones discussed in the previous subsection to facilitate their use~\cite{2018A&A...609A.130L,2020ApJ...905..102L, 2022ApJ...929..140Y}.

\section{Prospects}
\label{sec:prospects}
From the discussion in the two previous sections, it is evident that LIM serves both as a means of gathering statistical information about the Universe at high redshifts,  
and as a statistical probe of astrophysical evolution throughout the history of the Universe. As such, its prospects are unique and far-reaching. While current preliminary detections and upper limits have limited constraining power, forthcoming LIM observations will be able to yield invaluable information about galaxy formation and evolution, 
cosmology and fundamental physics. Here we describe some of the most motivated proposals to fulfill these goals, discussing the inherent sensitivity of LIM to the relevant signatures.

\subsection{Astrophysics}
\label{sec:Astro}

\vspace{-0.025in}

As emission from all sources is aggregated in  line-intensity maps, component separation and information extraction become challenging and highly model-dependent.
However, there is hope to mitigate this limitation by improving our understanding of the processes triggering different line emissions through line cross-correlations and combinations with external observables, as well as by comparing with targeted detailed observations of selected samples (see Sec.~\ref{sec:crosscorrelation}). The prospects for LIM to deepen our knowledge about astrophysics are mostly based on improving our understanding of the connection between the IGM and ISM properties and the line intensities, discussed in Sec.~\ref{sec:modeling}. Below we (somewhat artificially) distinguish between the potential of LIM to probe the star-formation history, the properties of the ISM, and the process of reionization.

\vspace{-0.075in}

\subsubsection{Star-formation history}

\vspace{-0.025in}

Star-formation rates are a proxy for the stellar emission of young stars, which are one of the main drivers of galaxy evolution, as well as for determining its chemical evolution  and gas reservoirs (in a circular, self-regulated, process~\cite{2011ARA&A..49..373B,Fu:2012qt,Carilli:2013qm}). 
Most of our current understanding of the high-redshift star-formation rate comes from optical and UV observations of stellar light and emission lines from the hot ionized gas in the ISM, but around half of the starlight is re-processed by dust~\cite{Casey:2018hlz} and only the brightest galaxies can be detected by galaxy surveys~\cite{Sargent:2013sxa, Kistler:2013jza, 2017MNRAS.467.1222H}, which leads to great uncertainties. Observations and analytic studies support a universal, power-law relation between the star-formation rate and gas content of the galaxy~\cite{Kennicutt:1998zb, Leroy:2008kh,Daddi:2010nn,Fu:2010qc, Krumholz:2011jm, 2015ApJ...805...31L, 2019ApJ...872...16D, 2021ApJ...908...61K}, depending on stellar feedback and other astrophysical processes~\cite{2019MNRAS.488.4753D}, but its potential redshift evolution is unknown~\cite{Santini:2013yfa, 2021ApJ...908...61K}.  
LIM can trace: (i) the cold molecular gas where stars form using CO lines~\cite{Righi:2008br,Mashian:2015his,Breysse:2016szq, Breysse:2016opl, Breysse:2015saa, Li:2015gqa} (with isotopologue line ratios  sensitive to the initial mass function~\cite{2019ApJ...879...17B}); (ii) the actual instantaneous star formation with optical hydrogen and oxygen lines~\cite{2013ApJ...763..132S, 2017ApJ...835..273G, 2018MNRAS.475.1587S}; (iii) the impact of star formation on the surrounding gas with fine-structure lines~\cite{Silva:2014ira, Yue:2015sua}; (iv) and the redshifted radiation from Pop III stars with helium and hydrogen lines~\cite{Visbal:2015sca, Parsons:2021qyw, Sun:2021drn}. All this, while surveying large enough volumes to provide ample statistics. 

Finally, Ref.~\cite{Sun:2022qrd} proposes to constrain the global star-formation law from the amplitude of a scale-dependent bias sourced by baryon fluctuations on baryon acoustic oscillations (BAO) scales~\cite{Barkana:2010zq,Angulo:2013qp,Schmidt:2016coo,Soumagnac:2016bjk} (still to be detected in galaxy clustering~\cite{Soumagnac:2016bjk,Soumagnac:2018atx}). The bias depends on the line luminosity (see Sec.~\ref{sec:formalism}), hence the target contribution can be isolated from the ratio of the power spectra of two lines. Non-linear clustering, which affects the BAO amplitude and the bias terms (see Ref.~\cite{Chen:2020ckc}), makes this approach challenging.

\subsubsection{The properties of the ISM}
The general properties of the ISM are expected to evolve substantially with redshift. The first galaxies are thought to be small, compact, both metal and dust poor, with a young stellar population, undergoing frequent mergers, exposed to a much weaker background radiation, etc. Moreover, early-galaxy populations are expected to be significantly less homogeneous, requiring larger samples to study global quantities. 

As stated above, LIM can probe many different scales and phases in the ISM and IGM thanks to its access to multiple emission lines and the sensitivity to all emitters. Molecular gas can be probed with the CO lines, especially combining CO isotopologues to obtain a better picture of high-density  clouds~\cite{Breysse:2016opl, 2018Natur.558..260Z}, and potentially using rotational transitions of hydrogen deuteride to target the earliest, ultra-low-metallicity galaxies~\cite{Bromm:2013iya}. Since the total gas mass is indirectly constrained by CIB measurements, combining HI (to separate between neutral and molecular gas abundances) and CO  reduces the uncertainties in the molecular gas mass to CO luminosity relation, though uncertainties dependent on gas temperature and metallicity remain~\cite{Blain:2002ec,2019ApJ...887..142S}. 
These degeneracies are further reduced by adding [CII] intensity maps to the analysis, due to their indirect dependence on the abundance of molecular gas. 

Meanwhile, the HII regions can be studied with the [NII] lines and their ratios, which are sensitive to the electron density and temperature~\cite{2019ApJ...887..142S, 2015ApJ...814..133G,2017ApJ...846...32D}. At the same time, Ly$\alpha$ can probe the circumgalactic medium, especially its ionization state and the gas density distribution~\cite{2013ApJ...763..132S, Pullen:2013dir, Comaschi:2015waa}, 
with improved sensitivity if cross-correlated with [CII]~\cite{Comaschi:2016soe}. 

LIM excels in probing the properties of the ISM and IGM in two more aspects. First, it is possible to study the ISM and IGM of specific galaxy populations by cross-correlating line-intensity maps with the corresponding sub-sample of galaxies under consideration~\cite{2019MNRAS.490..260B}. Secondly, most of the baryon content in the Universe is spread throughout  warm low density gas regions~\cite{Fukugita:1997bi, Cen:1998hc}; while this gas is undetectable by galaxy surveys, it can be probed with LIM, complementing other techniques such as the  Sunyaev-Zel'dovich effect~\cite{ACTPol:2015teu, deGraaff:2017byg}, as we discuss in Sec.~\ref{sec:crosscorrelation}.

\subsubsection{The process of reionization}

Reionization is the last phase transition the Universe has undergone, yet the EoR remains mostly unexplored by direct observations. This  is when the first stars and galaxies formed, and together with accreting black holes, gradually ionized the neutral gas surrounding them~\cite{Barkana:2000fd, 2013fgu..book.....L, 2016ARA&A..54..761S}. A simplistic view of the epoch of reionization depicts the IGM as a two-phase fluid, with `bubbles' of ionized gas growing around the first luminous sources with neutral gas filling the space in-between. Hence, the discussion of the previous subsection also applies to the study of reionization, when framed for lower metallicities and younger galaxies.

Star-formation and IGM models are very sensitive to assumptions related to the ISM  and the whole population of star-forming galaxies~\cite{2011ARA&A..49..373B}. This population can be efficiently probed with intensity maps of emission lines related to star formation~\cite{Gong:2011mf, Padmanabhan:2018yul}. In turn, Ly$\alpha$ probes some combination of the sources and the IGM~\cite{Pullen:2013dir,2013ApJ...763..132S,Visbal:2018dsi}. Thus, cross-correlation between these lines or with HI, which traces the neutral gas, will significantly improve our ability to map and understand reionization and the interplay between ionizing sources and the IGM~\cite{Lidz:2008ry,Gong:2011mf,Lidz:2011dx, Dumitru:2018tgh, Padmanabhan:2021tjr, Cox:2022hxl}, mitigating limitations due to foregrounds~\cite{Lidz:2011dx} and assumptions about escape fractions for ionizing photons~\cite{2011ApJ...730...48P}. Finally, to probe the evolution of reionization, one can use the anti-symmetric cross-correlations of HI and other lines~\cite{Sato-Polito:2020qpc, Zhou:2020hqh, Zhou:2020woq}, which evolve oppositely as reionization progresses.

\subsection{Cosmology}
\label{sec:Cosmo}
It is important to place the discussion of the prospects for cosmology with LIM experiments targeting emission lines related with star formation in the suitable context. LIM, especially for these lines, is still in the pathfinder stage, with current-generation experiments probing small volumes, as shown in Fig.~\ref{fig:EXP}. Hence, it is evident that LIM needs the time to evolve and mature before it can be competitive with flagship CMB experiments such as CMB-S4~\cite{CMB-S4:2016ple} and galaxy surveys such as Euclid~\cite{Amendola:2016saw} and DESI~\cite{DESI:2016fyo}. 

Nevertheless, even if cosmology continues to be dominated by the CMB and galaxy surveys for the near future, LIM observations grant access to regimes and scales that are out of reach for other  observables; examples include access to redshifts in-between volumes probed by galaxy surveys and CMB experiments, and the sensitivity to the integrated emission from the faintest astrophysical sources in the Universe, which form in less massive collapsed objects than those traced by galaxy surveys. This is what makes LIM a unique and complementary cosmological probe with great promise to constrain physics beyond $\Lambda$CDM. In addition, combining multi-line observations over the same redshift volumes and using cross-correlations with other observables can be used to mitigate cosmic variance via the multi-tracer technique~\cite{Seljak:2008xr, McDonald:2008sh}.

Below we discuss the intrinsic benefits that the LIM particularities offer for cosmology. While some of the target signatures may be degenerate with astrophysics,
actual line intensities will affect LIM's potential through their impact on the signal-to-noise ratio of the measurements. Quantitative forecasts (under a given set of assumptions) can be found in Ref.~\cite{Karkare:2022bai}.

\subsubsection{Dark matter}
LIM can probe small-scale matter clustering in two ways: through small-scale clustering measurements and through its dependence on the abundance of collapsed objects. Hence, LIM is sensitive to the effects of dark matter models that change the statistics of biased tracers of small-scale dark matter fluctuations, such as self-interacting dark matter~\cite{Tulin:2017ara}, models of dark matter-baryon scattering~\cite{Dvorkin:2013cea}, and ultra-light axion dark matter~\cite{Hlozek:2014lca}. The high sensitivity of LIM to models that modify the halo mass function has been demonstrated for observations before reionization~\cite{Munoz:2019hjh, Jones:2021mrs, Sarkar:2022dvl} and at lower redshifts~\cite{Bauer:2020zsj, Libanore:2022ntl, Sabla2022}.

As it collects all incoming photons, LIM is naturally suitable for searches of exotic sources of radiation. Furthermore, LIM's spectral resolution provides additional information on the spectral energy distribution of this radiation, allowing to separate between different processes. In the case of dark matter radiative decays, the resulting radiation effectively produces an emission line that will show up in LIM experiments as an interloper~\cite{Creque-Sarbinowski:2018ebl}. Techniques proposed to model and deal with line-interloper contamination (see Sec.~\ref{sec:formalism}) can be adapted to target and maximize the signal from these decays~\cite{Bernal:2020lkd, Shirasaki:2021yrp}. 
For example, forecasts indicate that LIM will be one of the most sensitive probes of
a possible coupling between electron-volt-scale axions and photons~\cite{Adams:2022pbo}, which may weigh in on potential explanations~\cite{Bernal:2022wsu} for the recent excess measurement of the cosmic optical background~\cite{Lauer:2022fgc}. Similar strategies can be applied to look for neutrino decays~\cite{Bernal:2021ylz}, the detection of which would hint at physics beyond the standard model of particle physics. 

\subsubsection{Light relics}
Neutrinos act like free-streaming particles and cannot be confined to regions smaller than their free-streaming scale. 
This scale grows until 
neutrinos become non relativistic ($z\sim 100-200$), after which it begins to shrink. Thus, neutrinos suppress matter clustering 
in a scale- and time-dependent manner~\cite{Lesgourgues:2013sjj}. 

LIM  will yield clustering measurements across a very wide redshift range, complementing those from galaxy surveys at lower redshifts and tracking the redshift evolution of the neutrino-induced scale-dependent suppression. Tracking the evolution of this dependence can break parameter degeneracies between the CMB and large-scale structure probes~\cite{Yu:2018tem}, especially between the sum of neutrino masses and the amplitude of the primordial power spectrum, the CMB optical depth to reionization, and the dark energy equation of state~\cite{Hannestad:2005gj,Liu:2015txa,Allison:2015qca}. LIM measurements split into several redshift bins will be highly sensitive to the neutrino mass~\cite{Bernal:2019jdo,MoradinezhadDizgah:2021upg}. Similar promise can be expected for other models involving light relics, such as dark matter decaying into lighter dark matter particles~\cite{FrancoAbellan:2021sxk}. Finally, LIM can contribute to constraining  $N_{\rm eff}$, the number of effective relativistic degrees of freedom in the early Universe, by extending searches for changes in the BAO phase~\cite{Baumann:2017lmt,Baumann:2019keh} to higher redshifts and larger volumes.

\subsubsection{Dark energy} 
The cosmic expansion history is strongly constrained at $z\lesssim 2.5$ from type-Ia supernovae~\cite{Brout:2022vxf} and BAO measurements from galaxy surveys~\cite{eBOSS:2020yzd}. At earlier times, we rely on extrapolations of an expansion determined by a matter-dominated Universe, which reproduce  existing measurements~\cite{Planck:2018vyg, ACT:2020gnv}. The lack of direct measurements may hide deviations from $\Lambda$CDM predictions that are degenerate with other parameters, such as ``tracking" dynamic dark-energy models, predicted by certain modified gravity theories (see e.g.\ Refs.~\cite{Raveri:2017qvt,Raveri:2019mxg}). 

The tomographic access to $z\gtrsim 2.5$ that LIM experiments provide will allow us to directly measure the expansion history of the Universe through BAO measurements~\cite{Chang:2007xk, Bernal:2019gfq, Karkare:2018sar}. While percent-level BAO measurements at $z>3$ will require stage-3 LIM experiments, current-generation experiments have the potential to return the first robust direct constraints on the expansion history at such redshifts (although anisotropies produced by line broadening may hinder the realization of the full potential of this cosmological probe~\cite{COMAP:2021rny}). 

LIM BAO measurements will bridge the gap between $H(z)$ constraints from galaxy surveys and CMB experiments, even for agnostic parameterizations~\cite{Bernal:2016gxb, Verde:2016ccp, Bernal:2021yli}, which will weigh in on the Hubble constant tension~\cite{Freedman:2021ahq, DiValentino:2021izs}. 
Moreover, pre-recombination deviations from $\Lambda$CDM proposed to solve the tension (see e.g.\ Refs.~\cite{Karwal:2016vyq,Poulin:2018cxd, Niedermann:2019olb}) affect the matter clustering also through perturbations of the new fields~\cite{Smith:2019ihp}. This significantly impacts the halo mass function, introducing deviations that increase with redshift~\cite{Klypin:2020tud}. LIM, which depends on the high-redshift halo mass function, will be especially sensitive to this effect.  Finally, tomographic growth-rate constraints with LIM clustering measurements~\cite{Bernal:2019jdo} will also improve constraints on modified gravity theories~\cite{Ferreira:2019xrr}.

\subsubsection{Inflation} 
The quintessential smoking gun for inflation is the detection of primordial gravitational waves, often parameterized by the tensor-to-scalar ratio. Currently, the most promising venue to detect such signatures is through primary $B$-modes in the CMB polarization~\cite{Kamionkowski:2015yta}. However, inflationary features may lie beyond the description of the tensor-to-scalar ratio~\cite{Achucarro:2022qrl}; some examples include 
primordial non-Gaussianity~\cite{Meerburg:2019qqi}, isocurvature perturbations~\cite{Bartolo:2001rt}, deviations from the almost scale-independent primordial power spectrum, spectral-index running~\cite{Munoz:2016owz}, power-spectrum oscillations~\cite{Zeng:2018ufm}, etc., all of which can be constrained  with large-scale structure measurements. Detecting or ruling out these features will provide invaluable information about inflation. As a leading example, a sizable local-type primordial non-Gaussianity can only be produced in scenarios with multi-field inflation~\cite{Maldacena:2002vr,Creminelli:2004yq,dePutter:2016trg}.

Primordial non-Gaussianity generically modifies the abundance of collapsed objects~\cite{Matarrese:2000iz}, induces scale dependence to the linear bias~\cite{Dalal:2007cu,Matarrese:2008nc}, and modifies the bispectrum~\cite{Creminelli:2003iq,Alishahiha:2004eh,Cheung:2007st,Senatore:2009gt,Chen:2006nt,Holman:2007na,Karagiannis:2019jjx}. These effects  are usually apparent at large scales and grow with redshift, hence making LIM a very promising way to improve current sensitivities~\cite{MoradinezhadDizgah:2018zrs,MoradinezhadDizgah:2018lac,Bernal:2019jdo,MoradinezhadDizgah:2020whw,Liu:2020izx,Chen:2021ykb}. Furthermore, the possibility to study and cross-correlate several lines in tomography up to high redshifts will help break degeneracies with compensated isocurvature perturbations~\cite{Sato-Polito:2020cil}, bias uncertainties~\cite{Barreira:2022sey}, and to reduce the limitations related with cosmic variance~\cite{Oxholm:2021zxp,Seljak:2008xr,McDonald:2008sh,Fonseca:2015laa,Alonso:2015sfa}.

\section{Formalism}
\label{sec:formalism}
LIM fluctuations are a biased tracer of matter density perturbations, but also depend on various astrophysical processes, which introduces additional non-Gaussianity in the line-intensity maps with respect to other tracers of the large scale structure. With the objective of recovering as much information as possible from line-intensity maps, several summary statistics have been proposed to exploit LIM observations. Below we will focus on the power spectrum (the formalism of which can be extended to configuration space or to higher order statistics), and on the voxel intensity distribution. We refer the reader to more focused references for other summary statistics such as the mapping of Ly$\alpha$ polarization~\cite{Mas-Ribas:2020wkz}, or the lensing of line-intensity maps~\cite{Foreman:2018gnv, Maniyar:2021arp}. 

Going back to Eq.~\eqref{eq:def_IandT},  
we define the conversion factors $X_{\rm LI}$ and $X_{\rm LT}$ between intensity or temperature to luminosity density as $\rho_{\rm L} = I/X_{\rm LI} = T/X_{\rm LT}  $. Expressions below are valid for both conventions, with the appropriate factor.

\subsection{Power spectrum}
\label{sec:powerspectrum}
\subsubsection{Intrinsic signal}
The LIM power spectrum has two main components: the clustering part, which follows the matter power spectrum, and a shot noise contribution due to the sources of the line not being a continuous field. The simplest formulation of the anisotropic power spectrum in redshift space is
\begin{equation}
        P(k,\mu) =  \left[X_{\rm LI}\int{\rm d}M \frac{{\rm d}n}{{\rm d}M}Lb_{\rm h}F_{\rm rsd}(k,\mu)\right]^2P_{\rm m}(k) + X_{\rm LI}^2\int{\rm d}M \frac{{\rm d}n}{{\rm d}M}L^2\,,
        \label{eq:Pk}
\end{equation} 
where all quantities also depend on redshift (and halo mass $M$, with the exception of $X_{\rm LI}$), $\mu \equiv \boldsymbol{k}\cdot\boldsymbol{k}_\parallel/k^2$ is the cosine of the angle between the Fourier mode $\boldsymbol{k}$ and its component $\boldsymbol{k}_\parallel$ along the line of sight, $k=\lvert \boldsymbol{k}\lvert$, $b_{\rm h}$ is a mass-dependent linear halo bias relating halo-number and matter perturbations, $F_{\rm rsd}$ is a factor encoding the effect of redshift-space distortions, $P_{\rm m}(k)$ is the linear matter power spectrum, 
${\rm d}n/{\rm d}M$ is the halo mass function, and we have assumed Poissonian shot noise. Note that any eventual scatter in the relations determining the luminosity function $L(M)$ must be taken into account in the integrals over the halo mass functions above. 
The clustering and shot noise terms depend on the first and second moments of the luminosity distribution, respectively. This means that, although the shot noise power spectrum often yields a considerably higher detection significance than the clustering component (especially for small-volume surveys), a significant detection of both components is required to break degeneracies in the $L(M)$ relation~\cite{2019MNRAS.490.1928Y}. 

The $F_{\rm rsd}$ factor includes the Kaiser effect~\cite{Kaiser:1987qv}, relevant at large scales, and a function suppressing the power spectrum at scales below a characteristic scale related to the halo pairwise velocity dispersion $\sigma_{\rm pv}$, to empirically reproduce the fingers-of-God effect~\cite{Jackson:1971sky}. Using for instance a Lorentzian function,
\begin{equation}
    F_{\rm rsd}(k,\mu) = 
    \frac{1+ \frac{f \mu^2}{b_{\rm h}(M)}}{1+0.5\left(k\mu\sigma_{\rm pv}(M)\right)^2}\,, 
\label{eq:RSD_bias}
\end{equation}
where $f$ is the growth rate.

The linear power spectrum, Eq.~\eqref{eq:Pk}, does not provide a good prediction of LIM clustering at small scales. This regime can be better modeled using the halo model~\cite{Cooray:2002dia}
\begin{equation}
\begin{split}
        P(k,\mu)  =  & \left[\left(X_{\rm LI}\int{\rm d}M \frac{{\rm d}n}{{\rm d}M} Lb_{\rm h}F_{\rm rsd}(k,\mu)\mathcal{U}(k) \right)^2P_{\rm m}(k) +  \right. \\
         & + \left. X_{\rm LI}^2\int{\rm d}M \frac{{\rm d}n}{{\rm d}M} L^2\mathcal{U}^2(k)\right. \Bigg] + P_{\rm shot}\,,
\end{split}
\label{eq:Pk_halomodel}
\end{equation} 
where $\mathcal{U}$ is the Fourier transform of the density profile (here assumed spherical) of a halo of mass $M$, and $P_{\rm shot}$ is the shot noise power spectrum from Eq.~\eqref{eq:Pk}. The first and second terms in the square brackets correspond to the two-halo and one-halo terms, respectively. Further non-linearities, in the matter power spectrum and in the bias expansion, can be accounted for (see Ref.~\cite{MoradinezhadDizgah:2021dei} for an example using effective field theory in the context of CO and [CII] intensity mapping). Furthermore, the contributions of each of the galaxies within a halo can be modeled: the halo occupation distribution model~\cite{Kravtsov:2003sg} can be generalized to LIM fluctuations rescaling the line luminosity of central and satellite galaxies accordingly~\cite{2019ApJ...887..142S}. In this case, the one-halo term includes the correlations between satellite and central galaxies and an additional shot noise term related with the number of emitters per halo~\cite{Schaan:2021gzb}.

So far, we have assumed a Dirac-delta line profile. However, velocity dispersion of the gas from which the lines are emitted broadens the emission line. In practice, this effect is similar to the fingers-of-God, but related to velocity dispersion within a single halo rather than pairwise velocity dispersion. In this case, the rotation velocity greatly dominates over thermal velocities. Assuming a Gaussian line profile, the gas rotation velocity $v$ broadens the line, resulting in a full-width half maximum of $\theta^\nu_{\rm FWHM}=\nu v(M)/c$. Unless $v(M)/c\ll 1$ (in practice, smaller than the spectral resolving power of the experiment, see below) the line broadening must be taken into account.  
The physical scale corresponding to the standard deviation of the broadened line-profile is $\sigma_v(M)=v(M)(1+z)/{H(z)\sqrt{8\log2}}$. The effect of the line broadening in the power spectrum is similar to the spectral resolution limit, but in this case the suppression depends on the halo mass and affects differently the clustering and shot noise contributions~\cite{COMAP:2021rny}:
\begin{equation}
\begin{split}
        P_{\rm broad} = & \left[X_{\rm LI}\int{\rm d}M \frac{{\rm d}n}{{\rm d}M}Lb_{\rm h}F_{\rm rsd}e^{-k^2\sigma_v^2(M)\mu^2/2}\right]^2P_{\rm m}(k) +\\
        +& X_{\rm LI}^2\int{\rm d}M \frac{{\rm d}n}{{\rm d}M}L^2e^{-k^2\sigma_v^2(M)\mu^2}\,.
        \label{eq:Pk_broad}
\end{split}
\end{equation} 

The power spectrum statistic suffers from degeneracies between the effect of the cosmological parameters and the dependence on astrophysics, that persist even after optimal reparameterizations~\cite{Bernal:2019jdo}. These degeneracies can be broken to a large degree by using mildly non-linear scales in the LIM power spectrum~\cite{Castorina:2019zho, MoradinezhadDizgah:2021dei}, combining it with higher order statistics~\cite{Gil-Marin:2014pva, Sarkar:2019ojl}, employing phase statistics~\cite{Wolstenhulme:2014cla, Byun:2020hun}, or using external priors on the astrophysical parameters~\cite{COMAP:2018svn, 2021arXiv211105354S}. 

\subsubsection{Observational effects}
LIM observations measure intensities as function of frequencies and angular positions on the sky, but three-dimensional clustering measurements need spatial distances. This requires interpreting  observed frequencies as redshifts (introducing projection effects for interloper lines, see below) and using a fiducial cosmology to transform these into distances. Assuming a background expansion that differs from the \textit{true} one adds an artificial anisotropy in the map, which is known as the Alcock-Paczynski effect~\cite{Alcock:1979mp}. This effect, in conjunction with  the BAO standard ruler, is used to extract the expansion history of the Universe from clustering measurements~\cite{Eisenstein:2003qy, Blake:2003rh, Seo:2003pu} with great precision~\cite{eBOSS:2020yzd} in a model independent way~\cite{Carter:2019ulk, Bernal:2020vbb}.

The observed power spectrum is
also limited by the spectral and angular resolutions of the experiment and the volume probed. Generally, a Gaussian beam with full-width half maximum $\theta_{\rm FWHM}=1.22c/\nu_{\rm obs}D$ for a dish with diameter D, is assumed.\footnote{Note that this applies for observations that only use the auto-correlation of each antenna. The angular resolution for interferometric observations depends instead on the maximum baseline distance.} The angular and spectral resolutions correspond to physical distance scales transverse to and along the line of sight as
\begin{equation}
    \sigma_\perp = D_{\rm M}\frac{\theta_{\rm FWHM}}{\sqrt{8\log 2}}\,, \qquad \sigma_\parallel = \frac{c\delta\nu(1+z)}{H\nu_{\rm obs}}\,,
    \label{eq:sigma_perp_parallel}
\end{equation}
respectively, where $\delta\nu$ is the minimum separation in frequency considered.\footnote{The minimum frequency division used in an experiment may be determined by its spectral resolution or by choices related with systematic effects of the analysis and map making.} The resolution limit results in a smearing of the observed line-intensity map, which can be modeled with a convolution in configuration space over a window function $W_{\rm res}$. Noting that $W_{\rm res} = W_{\rm res}^\perp W_{\rm res}^\parallel$ we  have that, in Fourier space,
\begin{equation}
     W_{\rm res}^\perp=\exp\left\lbrace -k^2\sigma_\perp^2(1-\mu^2)\right\rbrace\quad {\rm and} \quad W_{\rm res}^\parallel=\begin{cases}
     & \exp\left\lbrace -k^2\sigma_\parallel^2\mu^2\right\rbrace\,, \\
     & {\rm sinc}\left(\frac{k\mu\sigma_\parallel}{2}\right)\,, 
     \end{cases}
 \label{eq:Wk_res}
 \end{equation} 
depending on whether single Gaussian channels are used or many of them are stuck for $\delta\nu$ not related with spectral resolution.

\begin{figure}[t]
 \begin{centering}
\includegraphics[width=\textwidth]{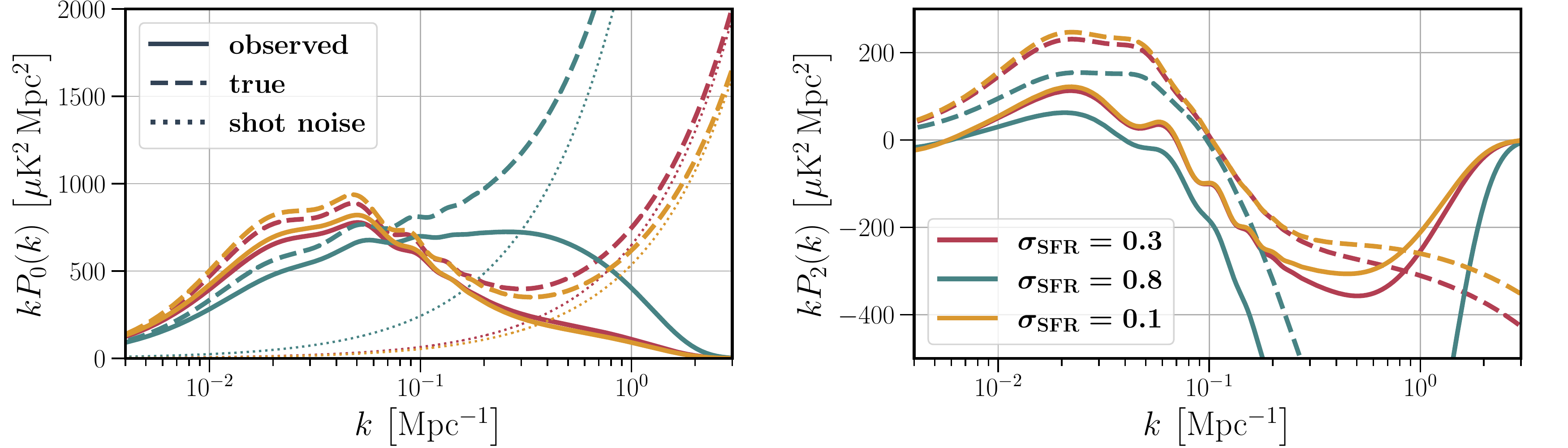}
\caption{CO power spectrum monopole (left) and quadrupole (right) at $z=2.9$. We show the true total power spectrum (dashed), the observed one (solid), and the shot noise power spectrum (dotted), varying the dispersion $\sigma_{\rm SFR}$ of the star formation rate to halo mass relation, assuming a beam of $\theta_{\rm FWHM}=1'$, spectral resolution of $\delta\nu=31.25$ MHz, and a cylindrical volume corresponding to 200 deg$^2$ and 7.7 GHz ($\Delta z=1$).}
\label{fig:Pk}
\end{centering}
\end{figure}

Similarly, the volume 
probed is determined by the total area $\Omega_{\rm field}$ of the sky surveyed and the frequency band of the experiment. We can model the loss of modes beyond the size of the observed volume and the effect of variable observing conditions with the mask $W_{\rm vol}$. The Fourier transform of the observed intensity fluctuations $\delta I(\pmb{x})$ is then
\begin{equation}
\begin{split}
    \delta\tilde{ I}(\pmb{k}) &= \int
    {\rm d}^3\pmb{x} e^{-i\pmb{k}\pmb{x}} W_{\rm vol}(\pmb{x})\int {\rm d}^3\pmb{x}'W_{\rm res}(\pmb{x}-\pmb{x}')\delta I(\pmb{x}') = \\& = \int \frac{{\rm d}^3\pmb{q}}{(2\pi)^3}  \delta I (\pmb{k}-\pmb{q}) W_{\rm res}(\pmb{k}-\pmb{q}) W_{\rm vol}(\pmb{q})\,,
\end{split}
\end{equation}
which results in an observed power spectrum
\begin{equation}
    \tilde{P}(\pmb{k}) = \int\frac{{\rm d}^3\pmb{q}}{(2\pi)^3} W^2_{\rm vol}(\pmb{q}) W^2_{\rm res}(\pmb{k}-\pmb{q})P(\pmb{k}-\pmb{q})\,,
\end{equation}
where the tilde denotes observed quantities. The anisotropic power spectrum cannot be measured, but it is possible to measure the Legendre multipoles directly, using e.g., the Yamamoto estimator~\cite{Yamamoto:2005dz}. The observed LIM power spectrum multipoles are~\cite{Chung:2019iim,Bernal:2019jdo}
\begin{equation}
\begin{split}
    \tilde{P}_{\ell}(k) = \frac{2\ell+1}{2}  \int_{-1}^{1}{\rm d}\mu \tilde{P}(k,\mu)\mathcal{L}_{\ell}(\mu),
\end{split}
\label{eq:multipole_scale}
\end{equation}
where $\mathcal{L}_{\ell}$ is the Legendre polynomial of degree $\ell$. In Fig.~\ref{fig:Pk}, we show the monopole and quadrupole of the intrinsic and observed power spectra, varying the scatter of the star-formation rate to halo mass relation.

\subsubsection{Covariance}
Experimental limitations also introduce an instrumental  white-noise floor in the LIM observations. The total noise variance per antenna per observed voxel when only the autocorrelation between antennas is considered is\footnote{The instrumental noise for interferometric experiments with simple configurations can be found in Ref.~\cite{Bull:2014rha}.} 
\begin{equation}
    \sigma_{{\rm N}, I}^2=\frac{\sigma_{\rm pix}^2}{N_{\rm feeds}N_{\rm pol}t_{\rm pix}}\,, \qquad   \sigma_{{\rm N},T}^2 = \frac{T_{\rm sys}^2}{N_{\rm feeds}N_{\rm pol}\delta\nu t_{\rm pix}}\,,
    \label{eq:sigmaN_T}
\end{equation}
for intensities and temperatures, respectively, where $\sigma_{\rm pix}$ is the noise equivalent intensity (NEI), $T_{\rm sys}$ is the instrument system temperature, $N_{\rm feeds}$ is the number of detectors per antenna, $N_{\rm pol}=1,2$ is the number of  polarizations the detectors are sensitive to, and $t_{\rm pix}$ is the observing time per pixel. One can also include the frequency dependence on the noise per voxel~\cite{Chung:2022lpr}.

The noise power spectrum is then  $P_{\rm N}= \sigma_{\rm N}^2V_{\rm vox}/N_{\rm ant}$, where $V_{\rm vox}$ is the volume of the voxel and $N_{\rm ant}$ the number of antennas used. 
Assuming Gaussianity and neglecting mode coupling, the power spectrum variance per $\mu$ and $k$ bin is 
 \begin{equation}
   \tilde{\sigma}^2(k,\mu)\equiv \left[\tilde{P}(k,\mu)+P_{\rm N}\right]^2/N_{\rm modes}\,,  
\label{eq:sigma2_tot}
 \end{equation}
where $N_{\rm modes} = V_{\rm field}k^2\Delta k\Delta \mu/(8\pi^2)$ is the number of modes observed per bin in a volume $V_{\rm field}$. Small volumes may be affected by higher cosmic variance due to limited Poisson sampling from luminosity functions~\cite{2020ApJ...904..127K}. The total covariance matrix for the power-spectrum multipoles is composed of the subcovariance matrices between $\ell$ and $\ell^\prime$ multipoles~\cite{Bernal:2019jdo}, 
\begin{equation}    
        \tilde{C}_{\ell\ell^\prime}(k)  = \frac{\left(2\ell +1\right)\left(2\ell^\prime +1 \right)}{2}\times \int_{-1}^{1}{\rm d}\mu \tilde{\sigma}^2(k,\mu)\mathcal{L}_\ell(\mu)\mathcal{L}_{\ell^\prime}(\mu).
    \label{eq:covariance}
\end{equation}

\subsubsection{Cross correlation}
As discussed in previous sections, cross-correlations of different lines reduce the impact of contaminants and allow for more detailed astrophysical analyses. Furthermore, the cross-correlations of three lines may be used to reconstruct their auto-power spectra free of contaminants~\cite{Beane:2018dzk}. The cross-power spectrum between  two lines can be computed following the formalism above, substituting all quadratic terms by the product of the contribution from each line and including their correlation coefficient~\cite{Pullen:2012su,Wolz:2017rlw,2019MNRAS.490..260B,2019ApJ...887..142S, Liu:2020izx}. At large enough scales, intensity fluctuations are completely correlated. Astrophysical dependence results in different $L(M)$ relations, which changes the line bias and may affect the correlation coefficient at intermediate and small scales. Thus, the correlation coefficient between  two lines is generally scale dependent.  

One effective way to formalize the inter-dependence between line luminosities and the host halo is through conditional luminosity functions~\cite{Yang:2002ww}, as proposed in Ref.~\cite{Schaan:2021gzb}. The conditional luminosity function models the mean number of galaxies in a halo of mass $M$ with luminosities $L_1,\dots, L_p$ for lines $1,\dots, p$. A complete modeling of the conditional luminosity function applied to the halo model self-consistently returns the one-halo term and shot noise contributions and accounts for any scale dependence of the correlation coefficient~\cite{Schaan:2021gzb}. However, modeling the conditional luminosity function is very challenging since, as discussed in previous sections, the modeling of each independent line at this point is already quite uncertain. 

The variance per $\mu$ and $k$ bin of the cross-power spectrum of two line-intensity maps is
\begin{equation}
\tilde{\sigma}^2_{XY} = \frac{1}{2}\left(\frac{\tilde{P}_{XY}^2}{N_{\rm modes}} + \tilde{\sigma}_X\tilde{\sigma}_Y  \right),
\label{eq:sigma2_cross}
\end{equation}
where $\tilde{\sigma}_X$ and $\tilde{\sigma}_Y$ are computed following Eq.~\eqref{eq:sigma2_tot} for each of the lines. 

\subsubsection{Angular power spectrum}

The angular power spectrum neglects the line-of-sight information within a redshift bin, but may be useful in cases with low spectral resolutions or to cross-correlate signals from different redshifts. Moreover, it is measured in observed coordinates and directly considers a curved sky, which avoids theoretical systematics related with wide-angle effects for wide surveys~\cite{Szalay:1997cc,Raccanelli:2010hk,Castorina:2017inr,Yoo:2013zga}. 
Tomographic angular analyses can better model the frequency evolution of the noise and the telescope beam, and can incorporate the experience from CMB analyses, such as the pseudo-$C_\ell$ technique~\cite{Hivon:2001jp,Tristram:2004if}. 

The angular power spectrum of maps $X$ and $Y$ as function of harmonic multipole $\ell$ at redshift bins $z_i$ and $z_j$ is given by
\begin{equation}
\mathcal{C}_\ell^{X,Y}(z_i,z_j) = 4\pi\int\frac{{\rm d}k}{k}\Delta_\ell^{X,z_i}(k)\Delta_\ell^{Y,z_j}(k)P_{\rm prim}(k)\,,
\label{eq:Cls}
\end{equation}
where $P_{\rm prim}$ is the dimensionless primordial matter power spectrum and the transfer function $\Delta_\ell$, accounting for the angular resolution and ignoring observational masks, is 
\begin{equation}
\Delta_\ell^{X,z_i}(k) = \int {\rm d}z \mathcal{W}^X(z,z_i)\Delta_\ell^{X}(k,z_i)\exp\left\lbrace-\frac{\ell(\ell+1)(\theta_{\rm FWHM}^X)^2}{16\log 2}\right\rbrace\,,
\label{eq:angular_transfer}
\end{equation}
where $\mathcal{W}$ is a normalized function centered on $z_i$ to delimit the redshift bin and $\Delta_\ell^X$ includes all the contributions to the observed perturbations of tracer $X$; for instance, the linear intrinsic clustering contribution corresponds to $\langle I_X b_X\rangle \mathcal{T}j_\ell(k\chi)$, where $\mathcal{T}$ is the matter transfer function, and $j_\ell$ is the $\ell$-th order spherical Bessel function.\footnote{For other contributions in the context of galaxy positions, see Ref.~\cite{DiDio:2013bqa}.} The corresponding shot noise 
\begin{equation}
    C_\ell^{X,Y,{\rm shot}} = \int \frac{{\rm d}z}{\chi^2}\mathcal{W}^X(z,z_i)\mathcal{W}^Y(z,z_j)\delta_{ij}^{\rm K}\int{\rm d}M\frac{{\rm d}n}{{\rm d}M}L_X(M)L_Y(M)\,,
\end{equation}
where $\delta^{\rm K}$ is the Kronecker Delta, must also be included. The observed angular power spectrum $\tilde{\mathcal{C}}_\ell^{X,Y}$ also includes the contribution from the noise  $N_{\ell}^X = \sigma_{\rm N,T}^2\delta\nu/(N_{\rm ant}\Delta_i\nu)\delta_{XY}^{\rm K}\delta_{ij}^{\rm K} = \sigma_{\rm N,I}^2/(N_{\rm ant}\Delta_i\nu)\delta_{XY}^{\rm K}\delta_{ij}^{\rm K}$, where $\Delta_i\nu$ is the frequency width of the $i$-th redshift bin. The covariance between the angular power spectra observed over a fraction $f_{\rm sky}$ of the sky is then
\begin{equation}
{\rm Cov}\left[\mathcal{C}_{\ell,(i,j)}^{X,Y},\mathcal{C}_{\ell,(p,q)}^{X,Y}\right] = \frac{\tilde{\mathcal{C}}_{\ell,(ip)}^{X,X}\tilde{\mathcal{C}}_{\ell,(jq)}^{Y,Y}+\tilde{\mathcal{C}}_{\ell,(iq)}^{X,Y}\tilde{\mathcal{C}}_{\ell,(jp)}^{X,Y}}{(2\ell+1)f_{\rm sky}}\,,
\label{eq:angular_cov}
\end{equation}
where the subscripts in brackets denote the redshift bins. The effects of contaminants and mode coupling due to incomplete sky coverage can  also be modeled under this framework, following e.g.,~Ref.~\cite{Anderson:2022svu}.

Spherical Fourier-Bessel analyses combine the main benefits of the three-dimensional and the angular power spectra, at the expense of numerical complexity. Ref.~\cite{Liu:2016xzv} developed a framework to analyze LIM clustering in this basis, including also observational effects. 
 
\subsection{Voxel intensity distribution}
\label{sec:VID}
The voxel intensity distribution (VID), which is the histogram of measured intensities, is an estimator for the probability distribution function (PDF) $\mathcal{P}(I)$ of the line intensity within a voxel. The VID encodes information about the whole intensity distribution, or luminosity function, providing complementary information to the power spectrum (which only probes its mean and variance), as demonstrated in Ref.~\cite{COMAP:2018kem, Libanore:2022ntl}. The intensity in a voxel is the sum of the intensities emitted by each of the $N_{\rm e}$ emitters it contains. If the line broadening is large enough,  the contribution from an emitter may extend beyond the voxel, which can be accounted for with a window function~\cite{Thiele:2020rig,Bernal2022}. 

The PDF of astrophysical sources is the sum of the conditional probabilities of having a given number of emitters in the voxel contributing to a total 
intensity~\cite{Breysse:2016szq}: $\mathcal{P}_{\rm astro}(I)=\sum \mathcal{P}_{N_{\rm e}}(I)\mathcal{P}_{\rm e}(N_{\rm e})$, where $\mathcal{P}_{\rm e}(N_{\rm e})$ and $\mathcal{P}_{N_{\rm e}}(I)$ are the PDFs of the number of emitters within a voxel and their total intensity, respectively. If there is no emitter in a voxel, the astrophysical intensity is null: $\mathcal{P}_0(I)=\delta_D(I)$. On the other hand, since the intensity is additive~\cite{Breysse:2016szq}, this can be written as a convolution 
\begin{equation}
    \mathcal{P}_{N_{\rm e}}(I) = (\underbrace{\mathcal{P}_1* \dotsc *\mathcal{P}_1}_{N_{\rm e}})(I) \,, \qquad
    \mathcal{P}_1(I) = \frac{V_{\rm vox}}{\bar{n}X_{\rm LI}}\left.\frac{{\rm d} n}{{\rm d} L}	\right\lvert_{L=\rho_L(I)V_{\rm vox}}\, ,
\label{eq:PT}
\end{equation}
where $\bar{n}$ is the mean comoving number density of emitters and `$*$' is the convolution operator. 

The number of emitters in a voxel obeys a Poisson draw with its mean equal to the expected number of emitters in that specific voxel, which depends on clustering. Ref.~\cite{Breysse:2016szq} assumes that the emitter number count follows the matter distribution, approximated with a log-normal distribution~\cite{1991MNRAS.248....1C, Kayo:2001gu}. 

Alternatively, $\mathcal{P}(I)$ can be explicitly expressed as a conditional probability depending on the emitter overdensity field $\delta_{\rm e}^{\rm v}$ smoothed over a voxel~\cite{Sato-Polito:2022fkd}. In this case, $\mathcal{P}_{\rm e}(N_{\rm e})$ is a Poisson distribution with space-dependent mean $N(\pmb{x}) = \bar{N}\left[1+\delta_{\rm e}^{\rm v}(\pmb{x})\right]$. The average over realizations can be later performed by invoking the Ergodic hypothesis. Although these two derivations are equivalent under the same set of assumptions, the latter allows the derivation of an analytic covariance between the VID and the power spectrum~\cite{Sato-Polito:2022fkd}, which depends on the integrated bispectrum~\cite{Chiang:2014oga} of one power of the emitter overdensity and two of the intensity fluctuations. This analytic covariance is consistent 
with simulation-based covariances~\cite{COMAP:2018kem}. In general, the correlation between the two summary statistics cannot be neglected for low instrumental noises (achievable by the final stages of current-generation experiments), and it peaks for low intensities and small scales (until resolution or instrumental noise limits the measurement).

\begin{figure}[t]
 \begin{centering}
\includegraphics[width=\textwidth]{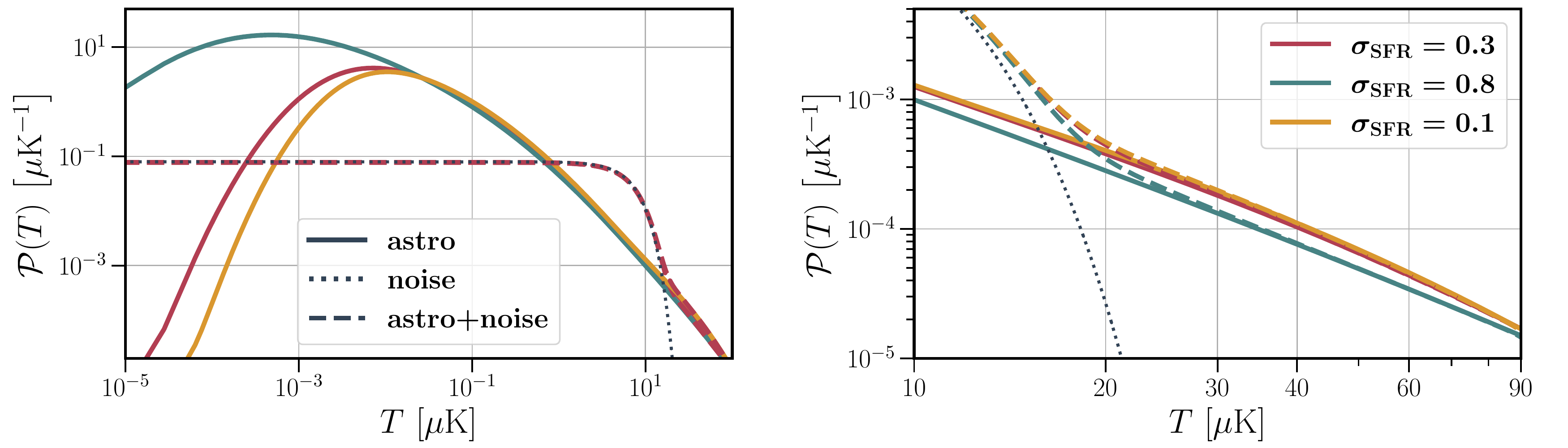}
\caption{CO temperature PDF for the same cases as in Fig.~\ref{fig:Pk}. We show the PDF from  the astrophysical contributions alone (solid), from the instrumental noise alone (dotted), and for their sum after mean subtraction (dashed); we assume $\sigma_N=5\,\mu{\rm K}$. 
The right panel zooms in on the tail of the PDF, which in practice is the only range from which it is possible to extract astrophysical information.}
\label{fig:VID}
\end{centering}
\end{figure}

The observed intensity also includes the contribution from instrumental noise, with a  PDF $\mathcal{P}_{\rm noise}$. The PDF for the total measured temperature in a voxel is then $\mathcal{P}_{\rm obs} = \left(\mathcal{P}_{\rm astro} * \mathcal{P}_{\rm noise}\right)(I)$. In the presence of foregrounds, or in case measuring absolute intensities is not possible, the ignorance about the zero point of the intensity PDF must be accounted for by considering only the PDF of the intensity perturbations: $\mathcal{P}_\Delta(\Delta I)\equiv \mathcal{P}(\Delta I+\langle{I}\rangle)$~\cite{Breysse:2016szq}. We show the astrophysical and total PDFs in Fig.~\ref{fig:VID}.

Finally, the predicted PDF must be connected to the observable, the VID 
$\mathcal{B}_i$. Using bins of width $\Delta I_i$, 
\begin{equation}
    \mathcal{B}_i = N_{\rm vox}\int_{I_i-\Delta I_i/2}^{I_i+\Delta I_i/2} \mathcal{P}_{\rm obs}(I){\rm d}I\,,
    \label{eq:VIDhist}
\end{equation}
where $N_{\rm vox}$ is the total number of voxels in the survey. To estimate the covariance, a first approximation assumes that $\mathcal{B}_i$ follows a multinomial distribution~\cite{Breysse:2016szq}, with negligible contributions from cosmic variance in most cases~\cite{Sato-Polito:2022fkd}. However, there is physical covariance between different intensity bins~\cite{COMAP:2018kem} that needs to be modeled, which requires a different formalism to compute the two-point PDF~\cite{Bernal2022}.

The VID can be useful in a variety of applications, besides direct parameter inference. For example, an extension of the VID formalism, in which the probability density distribution is conditioned on the number of detected galaxies in a voxel, has been proposed as a means to cancel out galactic foregrounds~\cite{Breysse:2019cdw}. Splitting the measured intensity into two contributions, $I=I_1+I_2$, where only $I_1$ is correlated with a field $y$, we can express the conditional PDF as
\begin{equation}
    \mathcal{P}(I\vert y)=\mathcal{P}(I_1\vert y)*\mathcal{P}(I_2\vert y) = \mathcal{P}(I_1\vert y)*\mathcal{P}(I_2)\,,
\end{equation}
since $\mathcal{P}(I_2\vert y)\equiv \mathcal{P}(I_2)$ by definition. Thus, the contributions from $I_2$ can be canceled out when comparing conditional PDFs for different values of $y$. An efficient way to do this in practice is with the ratio of the Fourier transform for different values of $y$. Taking $\mathcal{I}\equiv 2\pi/I$ as the Fourier conjugate of the intensity and $\breve{\mathcal{P}}(\mathcal{I}\vert y)$ as the Fourier transform of the conditional PDF,
\begin{equation}
    \frac{\breve{\mathcal{P}}(\mathcal{I}\vert y_1)}{\breve{\mathcal{P}}(\mathcal{I}\vert y_2)} = \frac{\breve{\mathcal{P}}(\mathcal{I}_1\vert y_1)\breve{\mathcal{P}}(\mathcal{I}_2)}{\breve{\mathcal{P}}(\mathcal{I}_1\vert y_2)\breve{\mathcal{P}}(\mathcal{I}_2)} = \frac{\breve{\mathcal{P}}(\mathcal{I}_1\vert y_1)}{\breve{\mathcal{P}}(\mathcal{I}_1\vert y_2)}\,,
\end{equation}
which is completely independent of the contributions of $I_2$. This ratio can be estimated using the ratio of conditional VIDs~\cite{Breysse:2019cdw}. 

The VID is complementary to the power spectrum beyond its access to the non-Gaussian information in the map. While the power spectrum is more sensitive to cosmology, only depending on astrophysics through integrals of the luminosity function, 
the VID depends directly on the (convolutions of the) luminosity function, and is only sensitive to cosmology through the expected number of galaxies per voxel. 
Hence, combining the power spectrum and the VID can help break the degeneracy between cosmology and astrophysics~\cite{COMAP:2018kem, Sato-Polito:2022fkd, Sabla2022}. 

\vspace{-0.1025in}

\subsection{Continuum emission}
\label{sec:continuum}
\vspace{-0.025in}
Intensity mapping of star-formation lines suffers from different observational contaminants than HI observations. At higher frequencies, Galactic foregrounds are less significant and better behaved~\cite{Planck:2016frx}. The main source of continuum foregrounds is the CIB. 
It traces the same objects that LIM targets, but is subject to different astrophysical processes and blends together all radial information~\cite{Shang:2011mh}. This contribution is mixed with the continuum emission of foreground galaxies. After its removal, residual continuum foreground contamination is mostly contained in the lowest line-of-sight Fourier modes, especially for experiments with high spectral resolution~\cite{Keating:2015qva, Switzer:2015ria, Switzer:2017kkz,Switzer:2018tel}, a limitation that may be partially lifted by applying neural networks~\cite{Pfeffer:2019pca, Moriwaki:2020bpr}.

Uncorrelated continuum emission (from the Milky Way and foreground galaxies) cancels in cross-correlations between LIM and galaxy surveys. However, the correlated continuum emission (from the actual galaxies that LIM traces) prevails, which enables the reconstruction of the galaxy spectral energy distribution, combining line and dust emission~\cite{Serra:2016jzs,Cheng:2021wex}.

\subsection{Line interlopers}
\label{sec:interlopers}
\vspace{-0.025in}
Line-interlopers can be challenging for LIM of star-formation lines, which are closer to each other in frequency. There are two main approaches to dealing with line interlopers: cleaning them from the observed maps, or modeling their contribution to avoid loss of signal and attempting to exploit the astrophysical and cosmological information encoded in their intensities.

Line-interlopers are usually sourced in different cosmic volumes than the target signal. Therefore, cross-correlating the map with an alternative tracer (e.g., galaxy surveys or the intensity of other spectral lines) of the cosmic volume from which the target signal comes leaves the interloper contribution out~\cite{Lidz:2008ry,Visbal:2010rz, Gong:2011mf, Chang:2015era, Silva:2014ira, Chang:2010jp, Masui:2012zc, BOSS:2015ids, Comaschi:2016pad, COMAP:2018svn, Visbal2022}. A similar approach can be used for the VID, using conditional distributions~\cite{Breysse:2019cdw, 2021arXiv211105354S}, as described above. These cross-correlations can later be used  to reconstruct the auto power spectra~\cite{Switzer:2013ewa, Beane:2018dzk}.

Another option involves the masking of voxels thought to be dominated by interloper contamination (including atmospheric lines for ground and balloon experiments). Masking can be either guided or blind. Guided masking~\cite{2018ApJ...856..107S, 2021arXiv211105354S} involves the use of external observations to identify interloper emitters. This approach cleans the interloper contribution from both bright and faint sources, minimizing the loss of information. Blind masking in turn can be used to clean the map in the absence of external observations. Assuming that the brightest voxels are likely dominated by the contribution from foreground interloper lines, as these typically come from lower redshift where galaxies tend to be more massive and more 
clustered, blind masking takes such voxels out to remove the brightest interloper contamination~\cite{Visbal:2011ee, Breysse:2015baa, Gong:2013xda}. Masking complicates the survey window and alters the bias and  amplitude of the shot noise, since it downsamples the intrinsic line-luminosity function, which hinders the theoretical interpretation of the observations~\cite{Visbal:2011ee}. 

The frequencies of the interlopers are known, which enables one to identify and characterize their contamination using spectral templates to fit the observed spectrum in each pixel map~\cite{Cheng:2020asz}. Knowledge about the interlopers can also be used to train deep learning models with simulations to separate components in the observed maps~\cite{2020MNRAS.496L..54M,Moriwaki:2020bpr, Moriwaki:2021yie}. Each method can reconstruct the VID and the power spectrum of each component, respectively, and both are limited by the instrumental noise, resolution, and size of the survey. 

\begin{figure}[t]
 \begin{centering}
\includegraphics[width=\textwidth]{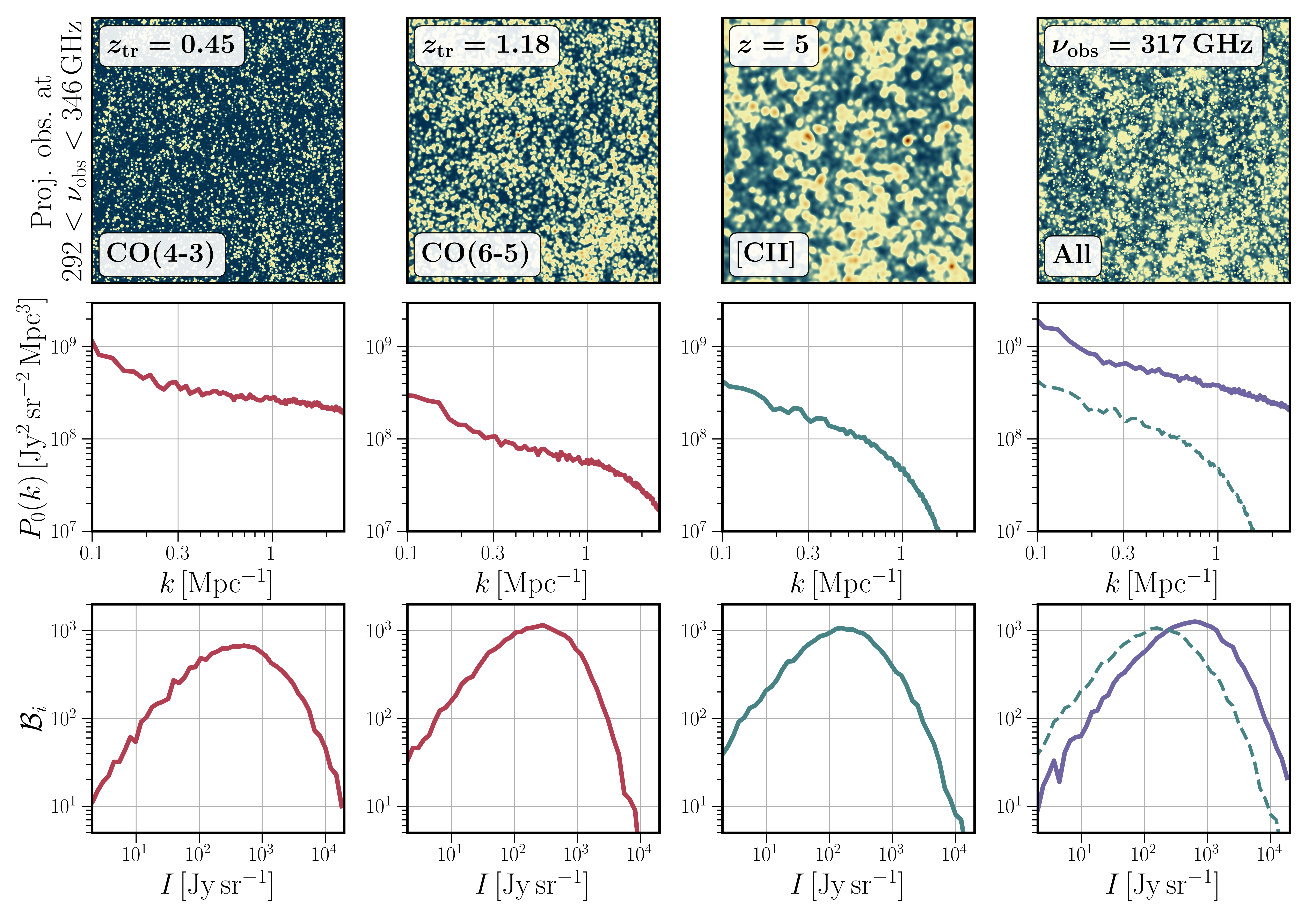}
\caption{Simulated noisless intensity fluctuations observed in the frequency range 292-346 GHz, targetting the [CII] line and including the projected contributions from CO(4-3) and CO(6-5) as interlopers, coming from $z=0.45$ and $1.18$, respectively, but with three-dimensional spatial scales ``wrongly" interpreted at $z=5$ (top panels). We also show the monopole of the power spectrum (middle panels) and the mean-subtracted VID (bottom panels). The three columns on the left show individual contributions from each line (CO lines in red, [CII] in blue), while the last column shows the sum of all contributions (in purple), with dotted lines showing the [CII]-only case to ease comparison. These results were obtained using the approach of Ref.~\cite{LC_paper}, assuming empirical relations from Refs.~\cite{2016ApJ...829...93K,DeLooze:2014dta}.}
\label{fig:interlopers}
\end{centering}
\end{figure}
 
Nevertheless, the contribution from line interlopers contains astrophysical and cosmological information. Together with the problems introduced by the methods discussed above, this motivates the modeling of the interloper contributions. The misestimation of the line redshift introduces projection effects: strong, artificial anisotropic distortions in the three-dimensional cubes~\cite{Gong:2013xda, Lidz:2016lub, 2017ApJ...835..273G, Cheng:2016yvu, Gong:2020lim}. Projection effects in clustering statistics can be modeled similarly to the Alcock-Paczynski (AP) effect, with a slight modification of the rescaling parameters:
\begin{equation}
    q_\perp=\frac{D_M(z_{\rm int})}{D_M(z_{\rm t})},\qquad q_\parallel = \frac{(1+z_{\rm int})/H(z_{\rm int})}{(1+z_{\rm t})/H(z_{\rm t})}\,,
\label{eq:scaling}
\end{equation}
where the subscripts `int' and `t' refer to the interloper and target line, respectively. Thus, the auto-power spectrum of the interloper contribution is $\tilde{P}_{\rm int}(k^{\rm infer}_\parallel,k^{\rm infer}_\perp) = \tilde{P}_{\rm int}(k_\parallel^{\rm meas}/q_\parallel,k_\perp^{\rm meas}/q_\perp)/(q_\parallel q_\perp^2)$, where beam smoothing, the mask and other filtering effects must be applied at $z_{\rm int}$. Assuming no overlap in the volume probed by each line, the total measured power spectrum multipoles are
\begin{equation}
    \tilde{P}_{\ell}^{\rm tot}(k^{\rm meas}) = \tilde{P}_\ell^{\rm t}(k^{\rm meas},z_{\rm t})+\sum\tilde{P}^{{\rm int}, i}_{\ell}(k^{\rm infer},z_{\rm int})\,.
\label{eq:Pelltot_interlopers}
\end{equation}
Note that the impact of the rescaling due to projection effects is usually significantly larger than the AP effect. The covariance of the total power spectrum is computed following Eq.~\eqref{eq:covariance} at $z_{\rm t}$, using $\tilde{P}_{\ell}^{\rm tot}$ to calculate $\tilde{\sigma}^2$.

The contribution from interlopers  also affects the measured VID, which can be modeled as an additional contribution to the PDF. In this case, the total measured intensity  is $I = I_{\rm t}+\sum I_{{\rm int}, i}+I_{\rm noise}$, hence the total VID is
\begin{equation}
    \mathcal{P}_{\rm tot+\chi}=\left(\mathcal{P}_{\rm t}*\mathcal{P}_{\rm noise}*\mathcal{P}_{\rm all\, int}\right)(I)\,,
\label{eq:Prob_InterLine}
\end{equation}
where $\mathcal{P}_{\rm all\, int}$ is the multiple convolution of the PDF for each interloper. We show an example of the contribution of interlopers to the power spectrum monopole and the VID in Fig.~\ref{fig:interlopers}. 

As mentioned above, the techniques to model known interlopers can also be used to look for exotic mono-energetic contributions to the intensity maps, such as radiative decays of dark matter and neutrinos~\cite{Bernal:2020lkd,Bernal:2021ylz}.

\subsection{Mocks}
\label{sec:mocks}
LIM modeling, the improvement of summary statistics to describe the map, and the development of techniques to deal with contaminants, all benefit from mock LIM observations. Contrary to analytic studies, mocks allow us to account for the whole distribution of galaxy properties and the scatter of scaling relationships. In addition, mocks account more accurately for non-linear clustering and allow for a more realistic inclusion of observational effects. For all these reasons, mocks are in practice the only way to fully account for the complexity and highly non-Gaussian nature of LIM, which has promoted their use as a gold standard in the field. 

In most cases,  LIM mocks are obtained by post-processing halo catalogs, a method also known as ``painting": line luminosities are assigned to halos based on scaling relationships calibrated on observations or theoretical models (see Sec.~\ref{sec:modeling}). The main benefits of this method are its fast application and the ability to consider any emission line and continuum radiation. 
The accuracy standards of the halo clustering (e.g.\ using approximate or exact cosmological N-body simulations) vary depending on the number of realizations required. Although there are examples using coeval boxes (snapshots of a simulation for which the whole volume is at the exact same redshift), it is important to build LIM mocks in lightcones in order to capture the evolution of not only the clustering of matter but also the ISM and IGM properties and their connections to line luminosities over the wide redshift intervals that LIM experiments often probe. 

The simplest approach assigns astrophysical properties to halos using external mean relationships~\cite{Silva:2014ira,Yue:2015sua,Li:2015gqa,COMAP:2018svn,COMAP:2018kem,2020MNRAS.496L..54M,MoradinezhadDizgah:2021dei,Murmu:2021quo, COMAP:2021rny}. Star-formation rates can be obtained from galaxy formation simulations~\cite{Vogelsberger:2014dza,Schaye:2014tpa,McCarthy:2016mry}, post-processed N-body simulations~\cite{DeLucia:2006szx,Guo:2010ap}, or empirical models based on abundance matching fit to observations~\cite{Behroozi:2012iw,Bethermin:2017ngy, Behroozi:2019kql}.  

Rather than using mean relationships, the halo-luminosity relation can be tracked more accurately if the halo astrophysical properties are evolved self-consistently. One option is to embed the abundance-matching results in the actual halo catalog that is painted~\cite{Chung:2018szp, Bethermin:2022lmd,LC_paper}; another involves the matching of semi-analytic models to  halo catalogues within a lightcone from N-body simulations~\cite{2021ApJ...911..132Y, 2022A&A...659A..12K}. Ref.~\cite{2021MNRAS.502.4858S} includes a comparison between semi-analytic models and the  UniverseMachine empirical forward model~\cite{Behroozi:2019kql}, which combines abundance-matching and a connection between star formation and halo accretion histories, yielding consistent results in the context of resolved astrophysical observations by the CANDELS survey~\cite{Grogin:2011ua, Koekemoer:2011ub}. Hence, LIM mocks using this combined approach are expected to yield similar results for consistent line models. 

Another approach is to extend the 21cmFAST~\cite{Mesinger:2007pd, Mesinger:2010ne}, which uses perturbation theory and excursion-set formalism to simulate HI LIM.  
A new code package, LIMFAST~\cite{Mas-Ribas:2022jok,Sun:2022ucx}, builds on 21cmFAST, adding galaxy-formation models, photoionization and stellar spectral energy distribution templates to  evolve the astrophysics of the ISM and IGM and compute the intensity and fluctuations of star-formation emission lines from the infrared to the UV.

Due to the numerical complexity and the small survey areas of future LIM experiments, most of the examples discussed in this sections are limited to very narrow lightcones. There are some exceptions, such as the simulated line intensity volume in a coeval box for CO and [CII] presented in Ref.~\cite{MoradinezhadDizgah:2021dei}. Furthermore, Ref.~\cite{LC_paper} presents a flexible pipeline to quickly generate LIM lightcone mocks for any line or experiment, accounting for line interlopers and continuum foregrounds, and offering the possibility to embed the resulting LIM mocks within a simulated sky 
including external observations such as radio galaxy surveys, CMB secondary anisotropies, etc.~\cite{Omori:2022uox}.

\section{Cross correlations}
\label{sec:crosscorrelation}

When LIM surveys overlap with other observables, they can be combined to boost sensitivity (and likely enable most of the first detections, see Sec.~\ref{sec:detections}), increase robustness against contaminants, reduce the impact of cosmic variance, and probe physics that individual observables are less sensitive to.

\begin{figure}[h!]
 \begin{centering}
\includegraphics[width=\textwidth]{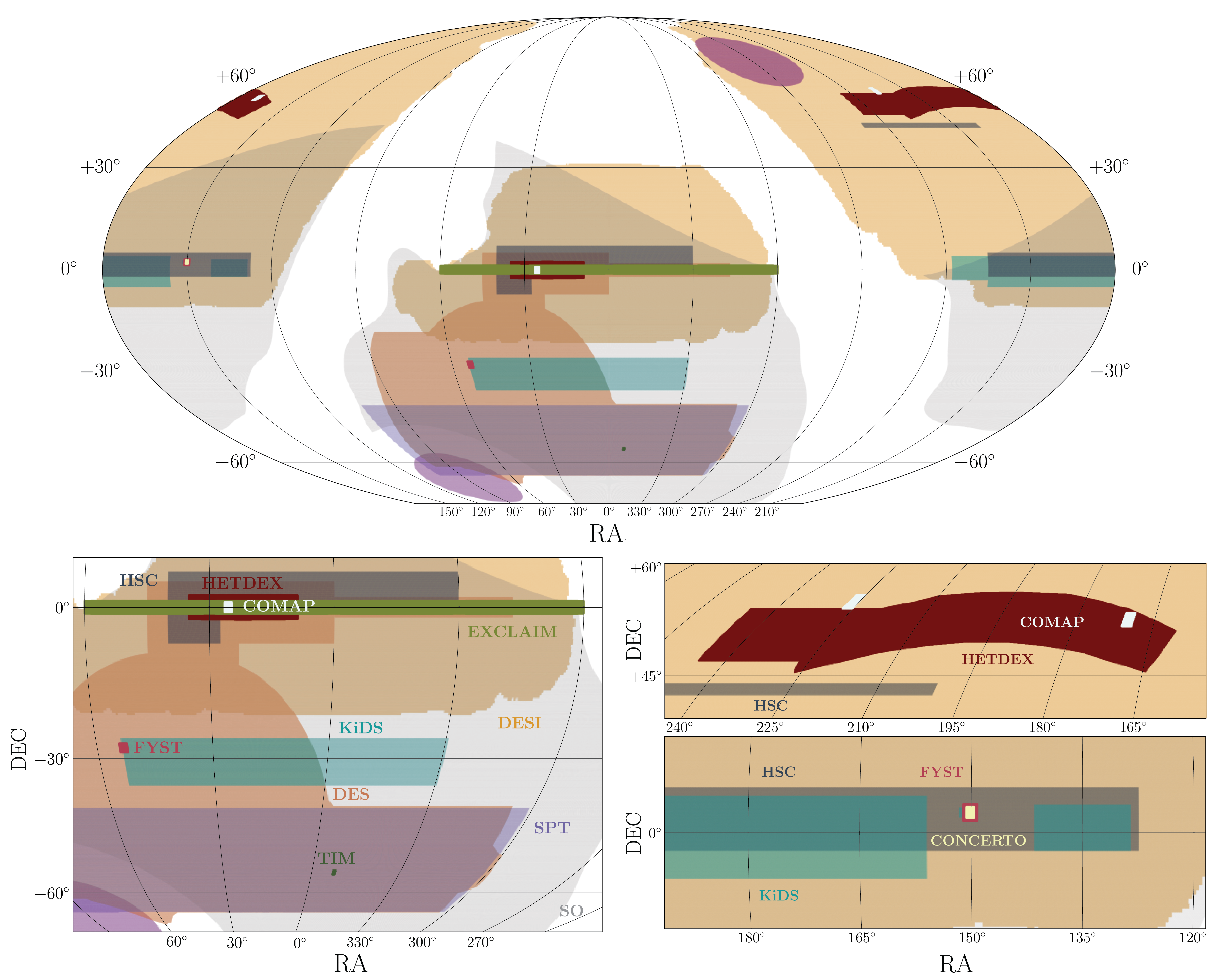}
\caption{Mollweide projection in equatorial coordinates of the observing fields of a selection of current and forthcoming LIM surveys, galaxy surveys (HSC, KiDS, DES and DESI), and CMB experiments (SPT and SO). Bottom panels show zoom-in regions with LIM surveys. The purple circular regions in the top panel correspond to SPHEREx deep fields (overlapping with Euclid deep fields). We do not show mmIME and COPSS due to their small footprints, nor SPT-SLIM and TIME, since they have not finalized their observing fields yet (SPT-SLIM will observe within the SPT field), nor COSMOS and CDFS fields as they coincide with FYST fields and have similar areas.}
\label{fig:sky}
\end{centering}
\end{figure}

In Fig.~\ref{fig:sky} we show an overview of the sky areas that will be observed by ongoing and forthcoming LIM experiments. Most LIM surveys either target fields around the celestial equator or prioritize existing multi-wavelength deep fields such as COSMOS (FYST, CONCERTO, mmIME); CDFS, GOODS-S and the Hubble ultra deep field (FYST, mmIME); and the Stripe-82 region (COMAP, EXCLAIM, HETDEX). Many LIM surveys will overlap on the sky, and some of them are part of experiments which also carry out a galaxy survey (as HETDEX and SPHEREx). Interestingly, all LIM surveys will overlap with at least one large galaxy survey\footnote{We do not show the Vera Rubin Observatory, which  targets the southern hemisphere, nor Euclid, which will survey the whole sky except for the Milky Way and  the Ecliptic.}, a wide HI LIM survey, and one CMB experiment (e.g.\ Planck, but many will also overlap with SPT, ACT or Simons Observatory). This makes line cross-correlations possible, with the benefits discussed in previous sections, as well as cross-correlations between LIM and other observables, the potential of which we detail further in this section.

\subsection{Galaxy surveys}
Line-intensity fluctuations are statistically correlated with other tracers of the large-scale structure, such as the galaxy distribution measured with galaxy surveys. 
Referring to quantities related to the galaxy field with a subscript `g', the linear cross-power spectrum of galaxy number counts and intensity fluctuations of a line $X$ is
\begin{equation}
P^{Xg} = \langle T_Xb_X\rangle  b_g F^{X}_{\rm RSD} F^{g}_{\rm RSD}P_{\rm m} +  X_{\rm LT}\frac{\langle \rho_L^X\rangle_g}{n_g}\,,
\label{eq:Pk_crossgal}
\end{equation}
where $b_g$ is the galaxy bias, $F^g_{\rm RSD}$ is defined using $b_g$, $\langle \rho_L^X\rangle_g$ is the mean luminosity-density of line $X$ sourced only from galaxies included in the galaxy catalog considered, $n_g$ is the number density of those galaxies, and we have assumed a Poissonian galaxy shot noise~\cite{Wolz:2017rlw,2019MNRAS.490..260B, 2021ApJ...915...33S}. Note that halo exclusion and clustering may cause galaxy shot noise to deviate from being Poissonian, changing its amplitude and even inducing a small scale dependence~\cite{Ginzburg:2017mgf,Schmittfull:2018yuk}. 
Since uncorrelated foregrounds and line interlopers come from a different volume than the galaxy survey, they are not correlated with the galaxy distribution and their contribution is left out of the cross correlation, as  discussed in Sec.~\ref{sec:interlopers}. Nonetheless, all components of the line-intensity map do contribute to the  cross-correlation covariance, which can be computed adapting Eq.~\eqref{eq:sigma2_cross}.

On the other hand, LIM-galaxy cross correlations can also benefit galaxy surveys. As galaxy surveys extend their reach to higher redshifts, spectroscopic surveys become harder and the redshift uncertainties in photometric and radio-continuum surveys become larger. Therefore, techniques like clustering-based redshifts~\cite{Menard:2013aaa,McQuinn:2013ib,Rahman:2014lfa}---which
uses a dataset with known redshifts to cross-correlate two-dimensional slices along the line of sight with the unknown dataset in order to retrieve the redshift distribution of the latter---can be applied using the superior spectral resolution of LIM surveys to improve the redshift determination of photometric galaxy surveys, as  has been demonstrated in the case of HI LIM~\cite{Alonso:2017dgh,Cunnington:2018zxg,Modi:2021okf}. This also enables tomographic analyses in radio-continuum surveys~\cite{Kovetz:2016hgp, Bernal:2018myq}.

In addition, cross correlations with specific galaxy populations can be used to obtain information about the IGM and ISM properties of such galaxies. At small scales, the cross-power spectrum depends on the mean line-intensity from that population of galaxies, which can be connected to the  line models discussed in Sec.~\ref{sec:modeling_theory}. Ref.~\cite{2019MNRAS.490..260B} cross-correlated quasar positions with CO LIM measurements to study the abundance and excitation state of molecular gas within galaxies hosting active galactic nuclei. This approach, complementary to stacking analyses, can be extended to any line or galaxy population.

Galaxy surveys also target gravitational weak lensing through cosmic shear and galaxy shapes. Cross correlating LIM with cosmic shear maps allows us to connect the intensity fluctuations with the actual matter density fluctuations~\cite{Schaan:2018yeh,Foreman:2018gnv,Schaan:2021gzb}. The lensing kernel can be considered using the following transfer function in Eq.~\eqref{eq:angular_transfer}:
\begin{equation}
    \Delta_\ell^\kappa(k,z) = \frac{3H_0^2\Omega_{\rm m}}{2c^2}(1+z)\int\displaylimits_z^{\rm z_{\rm max}}{\rm d}z'\frac{{\rm d}n_g}{{\rm d}z'}\frac{\chi(z)\left(\chi(z')-\chi(z)\right)}{\chi(z')}\mathcal{T}(k)j_\ell(k\chi)\,,
\end{equation}
where ${\rm d}n_g/{\rm d}z$ is the galaxy redshift distribution normalized to unity.  
Ref.~\cite{Chung:2022lpr} presents 
a systematic study of the prospects of cross correlating LIM  and projected cosmic shear maps, finding that current pathfinder experiments are expected to yield a marginally significant cross correlation with a survey like LSST, but future stage-3 experiments conceivably ready by the end of LSST will yield a high significance detection. 

Degeneracies between LIM astrophysical dependence and intrinsic alignments may weaken the astrophysical constraints from LIM-shear analyses, but they will strongly constrain integrated quantities like the cosmic density of molecular hydrogen or the star-formation rate. Pushing LIM-shear studies to smaller scales will shed light on the baryonic feedback on matter overdensities. 
Furthermore, LIM-shear correlations provide additional information to constrain radiative decays of dark matter~\cite{Shirasaki:2021yrp}. 

Potential improvements of this program include the use of tomographic cosmic shear, which is already available for shear auto-correlations and cross correlations with galaxy positions. However, using tomographic binning enhances the covariance due to shot noise, shape noise and the contribution from lines with null correlation with the cosmic shear bin. Finally, standard 3x2 analyses (galaxy clustering, galaxy clustering cross shear, and shear correlations)~\cite{Heymans:2020gsg, DES:2021wwk} could be extended to 6x2 analyses (galaxy clustering, galaxy clustering cross shear, galaxy clustering cross LIM, shear, shear cross LIM, and LIM correlations). 

\begin{figure}[h!]
\begin{centering}
\includegraphics[width=\textwidth]{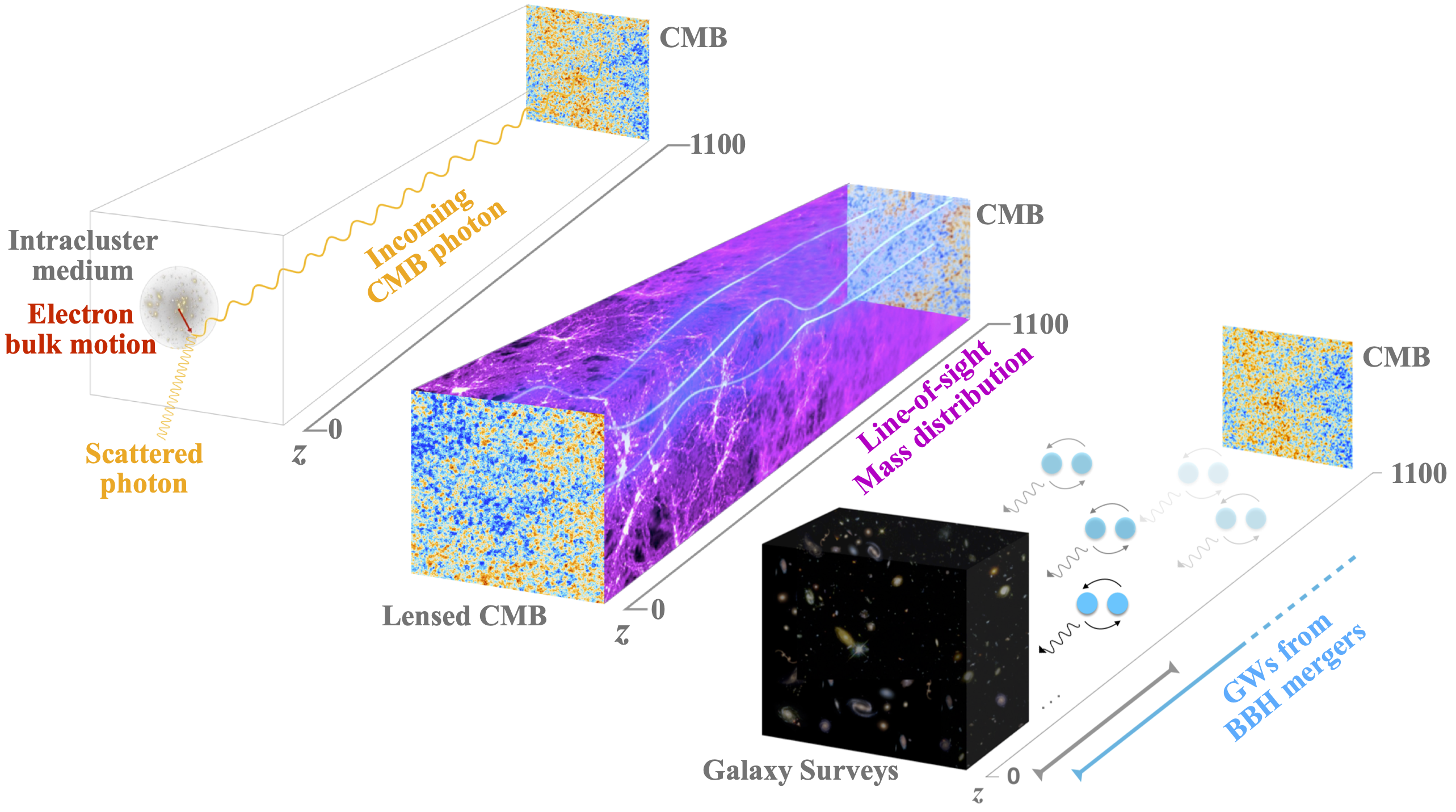}
\caption{LIM can be cross correlated with other tracers of the line-of-sight mass distribution, including CMB secondaries such as the kinetic Sunyaev-Zel'dovich effect or weak gravitational lensing and transients like gravitational waves from binary black hole mergers.}
\label{fig:LIMCrosses}
\end{centering}
\end{figure}

\subsection{CMB Secondary Anisotropies}

CMB photons are exposed to the effects of cosmological structures that they encounter along their path as they travel from the last scattering surface to us. These interactions inflict further perturbations in their temperature and polarization, known as secondary anisotropies~\cite{Aghanim:2007bt}, including gravitational effects such as lensing~\cite{Lewis:2006fu}, integrated Sachs-Wolfe effect~\cite{1967ApJ...147...73S, Rees:1968zza}, and moving lens effects~\cite{1983Natur.302..315B}, as well as photon-electron scattering described by different flavors of the Sunyaev-Zel'dovich (SZ) effect~\cite{Zeldovich:1969ff, Sunyaev:1970er, Sunyaev:1972eq, Sazonov:1999zp}. Figure~\ref{fig:LIMCrosses} illustrates some of these processes (as well as the overlap with transients, discussed later).

These secondary anisotropies can be detected in the CMB power spectrum, but are better isolated from the primary anisotropies in cross-correlation analyses with other tracers of large-scale structure. 
This allows for tomographic analyses instead of being limited to the integrated effect alone in CMB-only studies. This approach has been applied using galaxy surveys~\cite{White:2021yvw, DES:2018nlb, Krolewski:2021znk,  Hotinli:2018yyc, AtacamaCosmologyTelescope:2020wtv}; here we motivate the use of LIM in this context. 

First, LIM-CMB cross correlations enable the extension of CMB secondary anisotropies tomography to higher redshifts. This is motivated by the wide support of some CMB secondary kernels in redshift, which goes beyond the current reach of galaxy surveys. LIM can also be used to remove the contribution from secondary anisotropies in CMB observations. The main contaminant is gravitational lensing, but its effect can be partially reversed internally or using external observations~\cite{Smith:2010gu, Carron:2017vfg}, which improves cosmological parameter inference~\cite{SPTpol:2020rqg, BaleatoLizancos:2020mic, Hotinli:2021umk}. The CIB~\cite{Sherwin:2015baa, Larsen:2016wpa} and especially LIM~\cite{Karkare:2019qla} can be used to improve this procedure. Alternatively, in the same way that CMB lensing can be used to improve the modeling of the CIB~\cite{McCarthy:2020qjf}, it may be possible to improve the modeling of the line emission. 
Furthermore, LIM properties provide access to intrinsically different information, as is the case of probing reionization with the cross correlation of high-redshift HI and SZ maps~\cite{Ma:2017gey, Li:2018izh, LaPlante:2020nxx}. 

The late-time kinematic-SZ (kSZ) effect, detected both in the CMB temperature power spectrum~\cite{George:2014oba} and in cross correlation with galaxy surveys~\cite{Hand:2012ui,Planck:2015ywj,DES:2016umt,DeBernardis:2016pdv,AtacamaCosmologyTelescope:2020wtv}, can be used to reconstruct the matter velocity field at large scales. This reconstruction can be performed as a function of redshift by cross correlating kSZ maps with large-scale structure tracers~\cite{Ho:2009iw,Shao:2010md,Smith:2018bpn}. Ref.~\cite{Sato-Polito:2022fkd} presented a study of the potential to measure late-time kSZ tomography using LIM instead of galaxy surveys, with the [CII] emission line as an example. While a detection is beyond  reach for the initial design of FYST, an upgraded design with higher sensitivity can achieve a $\sim 3\sigma$ detection in cross correlation with Simons Observatory at $z\sim 3.7$, with substantial improvements for third-generation LIM surveys, yielding up to $\sim 10^2\!-\!10^3$ detection significance. 

The main limitation of CMB and LIM cross correlations is related to the contamination from continuum foregrounds. CMB secondary anisotropies are sensitive to quantities integrated along the line of sight, since they affect CMB photons as they propagate, which means that most of the information in their cross-correlation with large-scale structure tracers is contained in the long-wavelength line-of-sight components. As discussed in Sec.~\ref{sec:interlopers}, these are the most affected by continuum foregrounds, which has been shown to limit the potential of reionization kSZ-HI cross-correlations~\cite{LaPlante:2020nxx}. This limitation also applies, to a lesser degree, to other cross-correlations discussed in this section.  However, using high-order statistics or forward modelling~\cite{Modi:2019hnu} may retrieve some of the information.

\vspace{-0.125in}
\subsection{Transient Catalogs}
\vspace{-0.025in}

The recent discovery of gravitational-waves (GWs) from  mergers of binary black holes and neutron stars has kick-started, together with neutrino observations, the era of multi-messenger astronomy. Over the past few years, the LIGO-VIRGO-KAGRA collaboration~\cite{KAGRA:2013rdx} has detected almost $100$ merger events in three observing runs. With 
planned improvements towards the next observing runs and the advent of next-generation observatories such as Cosmic Explorer~\cite{LIGOScientific:2016wof} and the Einstein Telescope~\cite{Maggiore:2019uih},  future GW catalogs are expected
to amount to tens of thousands of events or more. As these mergers are detectable to high redshifts, GWs are destined to become a useful tracer of large-scale structure~\cite{Libanore:2020fim}. 

Several other classes of astrophysical transients  are within sight  of becoming useful standalone tracers of large-scale structure at high-redshifts. These include gamma-ray bursts, supernovae, tidal disruption events and fast radio bursts (FRBs). Current catalogs of all these types of transients  
contain between dozens to thousand of events, but with ongoing and upcoming experimental programs we can expect at least a ten-fold increase in number over a ten-year timescale. For FRBs in particular, due to their extremely high rate~\cite{Fialkov:2017qoz}, we may expect a few orders of magnitude more events in future catalogs from CHIME~\cite{CHIMEFRB:2021srp} and other instruments.

One of the main drawbacks of transient-event catalogs such as GWs and FRBs is their poor redshift determination. A promising method to curb this limitation is clustering-based redshift estimation, mentioned above. Here LIM has potential for dramatic payoff, as its redshift information is by definition controllable via the observation frequency and its two-dimensional clustering can be extracted up to very high redshifts (see Fig.~\ref{fig:epochs}). 
Reconstructed redshift distributions will allow tomography, which enables several cosmological applications. For example, it can be used to greatly improve parameter estimation~\cite{Kovetz:2016hgp}, to constrain the expansion history of the Universe (when comparing redshifts to the luminosity distance 
or dispersion measure for GWs or FRBs, respectively)~\cite{Scelfo:2021fqe,Mukherjee:2021bmw}, to measure the CMB optical depth to reionization~\cite{Fialkov:2016fjb}, to break the kinetic Sunyaev-Zel'dovich effect optical depth degeneracy~\cite{Madhavacheril:2019buy} and to constrain the gravitational-slip in modified gravity theories~\cite{Abadi:2021ysz}, to name a few.

Moreover, as both compact binary mergers and FRBs are expected to be related to properties of the stellar population, such as the star-formation rate and stellar mass in the host galaxy~\cite{Artale:2019tfl}, the  metallicity distribution~\cite{Dvorkin:2016wac}, etc.,  
there is promising complementarity between LIM and these transients. Their cross-correlation can help distinguish between GW progenitors~\cite{Raccanelli:2016cud}, constrain the origins of FRBs~\cite{Zhang:2020ass}, infer the delay-time distribution between binary formation and merger~\cite{Safarzadeh:2019pis}, and much more.

\vspace{-0.15in}
\section{Conclusions}
\label{sec:conclusions}

Here concludes our short review of the theory of LIM. After describing the basic idea behind LIM and summarizing the current experimental landscape, we introduced the primary emission lines targeted by these experiments, from the rotational CO transitions in the sub-mm, through the series of fine-structure lines such as [CII] in the infrared to hydrogen lines including H$\alpha$ and Ly$\alpha$ observed in the optical and UV, and discussed different approaches to modeling their luminosity functions. We then surveyed the prospects to use LIM to advance our understanding of major topics in astrophysics and cosmology, from star-formation history, the properties of the ISM and the process of reionization, to the search for dark matter and light relics and the study of cosmic inflation and dark energy. The heart of the review was dedicated to carefully laying out the modern formalism to describe the statistics of line-intensity maps, including the power spectrum and voxel-intensity distribution, with special attention to observational effects and cross-correlation. We also discussed the problem of interloper foregrounds which uniquely plagues this type of observation. Finally, we devoted a separate chapter to cross-correlation opportunities between LIM and other observables, from galaxy surveys to CMB secondary anisotropies to astrophysical transients. 

We hope that this review  successfully demonstrates the bright outlook for LIM as a burgeoning observable in high-redshift astrophysics and cosmology, that it provides a useful self-contained description of the theoretical backbone of the field for experts and newcomers alike, and that it affords the necessary tools for novices to pick up the gauntlet and join the effort to make LIM science a prosperous reality. In this spirit, we accompany its release with an updated online version of the lim\footnote{\MYhref{https://github.com/jl-bernal/lim}{https://github.com/jl-bernal/lim}} package, and provide a few tutorial Jupyter notebooks that can be used to reproduce the main figures above.

\begin{acknowledgements}
We thank Dongwoo Chung, Kirit Karkare, Guilaine Lagache, Olivier Dor\'e, Tzu-Ching Chang, Karto Keating, Joaquin Vieira, Anthony Pullen and Eiichiro Komatsu for providing information necessary to produce Fig.~\ref{fig:EXP}. We also thank Karto Keating for assistance with  the modeling used in Fig.~\ref{fig:kernels}, Selim Hotinli for help with the SO map in Fig.~\ref{fig:sky} and Reut Kovetz for help with Fig.~\ref{fig:epochs}. We are grateful to Donwgoo Chung, Adam Lidz, Maja Lujan Niemeyer, Anthony Pullen, Emmanuel Schaan and Eric Switzer for thoughtful comments on the manuscript. JLB is supported by the Allan C.\ and Dorothy H. Davis Fellowship.
EDK is supported by a faculty fellowship from the Azrieli Foundation.
\end{acknowledgements}

\bibliography{inspiresBibtexMay23.bib}
\bibliographystyle{utcaps}

\end{document}